\documentclass[9pt,twocolumn,twoside]{osajnl}
\usepackage{setspace}
\usepackage{enumitem}
\usepackage{amssymb}
\usepackage{amsmath}
\usepackage{textgreek}
\usepackage[flushleft]{threeparttable}
\usepackage{algorithm}
\usepackage{caption}
\usepackage{subcaption}
\usepackage{xurl}
\usepackage{hyperref}
\usepackage{mathtools}
\usepackage{pgfplots}
\pgfplotsset{compat=1.17}
\usepackage{listings}
\lstdefinestyle{interfaces}{
  float=tp,
  floatplacement=tbp,
}

\journal{jocn} 

\setboolean{shortarticle}{false}

\title{RIFL: A Reliable Link Layer Network Protocol for Data Center Communication}

\author[1,*]{Qianfeng Shen}
\author[2]{Jun Zheng}
\author[1]{Paul Chow}

\affil[1]{The Edward S. Rogers Sr. Department of
Electrical \& Computer Engineering, University of Toronto, 10 King’s College Road
Toronto, Ontario, M5S3G4, Canada}
\affil[2]{Alibaba Group US, 525 Almanor Ave, Sunnyvale, CA, 94085, USA}

\affil[*]{Corresponding author: qianfeng.shen@mail.utoronto.ca}

\begin{abstract}
More and more latency-sensitive services and applications are being deployed into the data center. Performance can be limited by the high latency of the network interconnect. Because the conventional network stack is designed not only for LAN, but also for WAN, it carries a great amount of redundancy that is not required in a data center network. This paper introduces the concept of a three-layer protocol stack that can fulfill the exact demands of data center network communications. The detailed design and implementation of the first layer of the stack, which we call RIFL, is presented. A novel low latency in-band hop-by-hop re-transmission protocol is proposed and adopted in RIFL, which guarantees lossless transmission in a data center environment. Experimental results show that RIFL achieves 110 nanoseconds point-to-point latency on 10-meter Active Optical Cables, at a line rate of 112 Gbps. RIFL is a multi-lane protocol with scalable throughput up to multi-hundred gigabits per second. It can be the enabler of low latency, high throughput, flexible, scalable, and lossless data center networks.
\end{abstract}

\setboolean{displaycopyright}{true}

\begin{document}

\maketitle

\section{Introduction} \label{Introduction}
Major data center services and applications such as remote direct memory access (RDMA), machine learning, and cloud storage demand the network interconnect to be low latency and lossless while preserving high bandwidth. Previous works, such as ~\cite{sidler2020strom,DCQCN}, demonstrate how the performance of applications in various fields can be drastically impacted by interconnect latency. It is important to realize that most of the technologies and concepts used in today's data center networks (DCNs) existed before the large-scale data centers of today were even imagined. For example, IP protocols were first defined in 1974~\cite{IP}, well before any massive data center was built. Today, with rapidly evolving technologies, it is time to explore new approaches for the DCN that are designed for the needs of today's data center.

The conventional TCP/IP stack is designed to work reliably not only in a local area network (LAN), but also in a wide area network (WAN). The physical properties of a LAN and a WAN are significantly different. Both bandwidth-wise and latency-wise~\cite{dragojevic2014farm}, TCP/IP and UDP/IP carry too much redundancy when used in a LAN. Considering the diameter of a data center server room is rarely more than 100 meters, a DCN is essentially a LAN. There should be a more efficient protocol stack that fulfills the exact needs of a DCN.

Nevertheless, protocols based on TCP/IP and UDP/IP~\cite{DCTCP,quic} still dominate the data center market. One of the most important reasons for cloud providers to use these protocols is that hardware changes would be required to both the end devices and the network switches to deploy a new protocol in a data center. Traditionally, the network switches and the NICs are all implemented using ASICs. It would take years to design, fabricate, test, and deploy the ASICs for a new protocol.

Compatibility with the established infrastructure and the barrier to developing new ASICs makes it extremely difficult to introduce major changes. However, it is still interesting to know what opportunities exist that might influence DCN infrastructure over time. The basis of our work is to build an experimental platform that enables us to explore what might be possible if we could start over, i.e., how would we build the DCN infrastructure starting with what we know is feasible today and not be constrained by any legacy requirements, either technical or business. In this paper, we will show what we can do by leveraging the capabilities of modern FPGAs.

Today, the number of high-speed transceivers is quickly increasing in modern FPGAs. Off-the-shelf FPGAs containing multiple QSFP28 ports are already available in the market ~\cite{QSFPboard}, showing that a flexible and economically efficient approach to redesigning DCNs starting from the very bottom layer of the protocol stack can be prototyped without needing new ASICs.

There are many network protocols apart from TCP/IP and UDP/IP. However, some of them~\cite{INKL,Aurora} are dedicated to the link layer, providing limited scalability and flexibility. Some of them are based on the Media-Independent Interface (MII)~\cite{micro} or UDP~\cite{quic}, and you cannot remove the redundancy carried with the conventional network stack. Others such as Infiniband~\cite{Infiniband} implement re-transmission in their Transport Layer. We will discuss its inefficiency in Section \ref{S2}.

To meet the exact demands of a DCN, we propose a new protocol stack as follows:
\paragraph{Layer 1: Link Layer} This layer is implemented immediately next to the transceivers. It is a combination of the data link layer (layer 2) and the physical layer (layer 1) in the OSI model. It should provide a line protocol with appropriate data packetization, channel bonding and clock compensation. Re-transmission should also be a part of this layer to resolve link-level data corruption. The benefits of implementing re-transmission at this layer is discussed in Section \ref{S2}. Beyond this layer, there should be no data corruption caused by link noise.
\paragraph{Layer 2: Network Layer} This layer should provide a low latency routing scheme that avoids using a centralized routing table. Switch initiated congestion control mechanisms should also be implemented in this layer. Beyond this layer, all the data transfers should be lossless. Anything sitting above this layer does not have to worry about checksums, re-transmission, or congestion at all.
\paragraph{Layer 3: Application Layer} This layer consists of two parts: hardware and software. The hardware serves as an accelerator for common DCN applications and services, e.g., a near-memory computing engine to reduce the round trips for RDMA. The software abstracts the usage of the hardware and provides the software programmer an easy-to-use user interface.

With this protocol stack, we envision a lossless network can be built. In our prototype, at its Layer 2 interface, this network can provide lossless links with less than 300 ns typical latency per hop with bandwidths beyond 100 Gbps.

\textbf{This paper focuses on the Link Layer design, named RIFL}. The Network Layer and the Application Layer designs will be the subject of our future work. The rest of this paper is organized as follows: Section \ref{S2} discusses the physical properties of a DCN and how they can be leveraged to build a more efficient Link Layer protocol. In Section \ref{S3}, we define the RIFL Frames. Section \ref{S4} introduces the RIFL protocols. Section \ref{S5} presents the hardware implementation of RIFL. Section \ref{S6} provides performance results. Section \ref{S7} discusses the related work and Section \ref{S8} concludes this work.

\section{Layer 1 - The Link Layer} \label{S2}
The goal of our Layer 1 is to provide a reliable Link Layer point-to-point protocol as a foundation for the higher layers. This layer should be low-latency, high-bandwidth and use minimal hardware resources. Reliability here means correcting any bit errors that occur during transmission across the link. With a reliable link, the higher layers need not be concerned with any data integrity issues resulting from the physical transmission.

In this section we cover the following topics: The development of our Layer 1 first requires the selection of the mechanism for error detection and correction. After selecting re-transmission, we show that it can work within the constraints of a DCN. After justifying hop-by-hop Link Layer re-transmission, we show that an additional property can be introduced. Finally, we explain why we can solely rely on negative acknowledgments (NACKs) as the re-transmission notifications in DCNs, and why doing so is critical for the efficiency. Given these justifications we can then develop the circuit for our protocol implementation.

We start by imposing the first constraint:
\begin{enumerate}[label=\Alph*]\bfseries
\item the distance between any two nodes within a DCN is less than 500 meters.
\end{enumerate}

\subsection{Forward Error Correction (FEC) vs Re-transmission}

There are two major approaches to eliminate the effect of data corruption caused by bit errors: FEC and re-transmission.

FEC is widely used in wireless and low-level wired communication. It requires the sender to send redundant data along with the payload. The redundant data, which is usually an error correction code (ECC), can be used to detect the errors in the payload as well as correct the errors.

Re-transmission requires redundant data as well. The redundant data is usually a checksum. However, the checksum is not used to correct the errors. Instead, it only needs to carry enough information to detect the errors in the payload. While sending data to the receiver, the sender keeps a copy of the most recent transmitted data. Once an error is detected, the receiver notifies the sender to resend the corrupted data.

While FEC detects and corrects the errors, the checksum only detects the errors. Consequently, for the same size of the payload, the size of the ECC used by FEC is much larger than the size of the checksum used by re-transmission, which means the bandwidth overhead for FEC is much larger than re-transmission. Moreover, because FEC usually involves large matrix multiplications, the typical latency overhead for FEC is much larger as well. Therefore, FEC is more suitable in situations where re-transmission is impossible or very expensive. E.g, in one-way communications such as radio networks or simplex links, or in any bidirectional communication that operates on a very high bit error ratio (BER).

In current DCNs, 100G Ethernet is slowly becoming the dominant interconnect technology~\cite{top500}. The commercially available QSFP28 cables used by 100G Ethernet can guarantee BERs better than 10\textsuperscript{-12} without using FEC. Under such a low BER, re-transmission is much more efficient than FEC. However, as the next generation cable technologies pursue higher throughput per lane, their associated BER can be significantly higher than 10\textsuperscript{-12}. Thus, for better compatibility with the future technologies, our BER constraint is:
\begin{enumerate}[label=\Alph*]\bfseries
\setcounter{enumi}{1}
\item the effective BER of the link that RIFL operates on must not exceed 10\textsuperscript{-7}.
\end{enumerate}
We set the minimal BER requirement as 10\textsuperscript{-7} because in our simulations, we found that in any link shorter than 500 meters with a BER better than 10\textsuperscript{-7}, re-tranmission can be done efficiently. Plus, a minimal BER of 10\textsuperscript{-7} means RIFL can work not only with the current popular cables, but also with any future physical links providing BERs better than 10\textsuperscript{-7}. For links whose BERs are worse than 10\textsuperscript{-7}, FEC must be incorporated to guarantee reliable transmissions. Otherwise, the bandwidth will be mainly occupied by re-transmissions instead of regular data transmissions. Nevertheless, even if FEC is used, RIFL still has advantages because it only needs a lightweight FEC code to improve BER to better than 10\textsuperscript{-7} while other protocols, such as Ethernet, require much lower post-FEC BERs~\cite{IEEE_Eth}. Even with FEC, they still cannot guarantee lossless transmissions.

To summarize, we choose re-transmission as the main error recovery method for RIFL. When BER is higher than 10\textsuperscript{-7}, FEC has to be applied to improve the BER so that constraint B can be satisfied.

\subsection{Re-transmission Efficiency vs Round Trip Time (RTT)}
To guarantee a lossless link, the re-transmission mechanism should be designed for the worst case. Because any Frame\footnote{\textbf{Frame}: the basic unit of data transmitted across the link. Any data is transmitted along the link by the means of one or multiple Frames.} being transmitted during the RTT may have errors, the size of the re-transmission buffer, denoted as S\textsubscript{retrans}, must be larger than the size of the data being transmitted during the largest RTT between the sender and the receiver, namely:
\begin{equation}\label{e0}
S_\textit{retrans} \geqslant \lambda_\textit{line} * \textit{RTT}
\end{equation}
where \textlambda\textsubscript{line} denotes the line rate. 


The larger the RTT is, the larger the re-transmission buffer needs to be. It is worth noting that when line rate is larger than 100 Gbps and RTT exceeds 100 $\mu s$, it requires more than one megabytes of re-transmission buffer. It is no longer suitable to use embedded memories such as SRAM as the buffer. Otherwise, the circuit area will be too large. This issue is encountered by some TCP implementations ~\cite{limago,sidler10G}. Their solution is to use DDR memory as an alternative. However, it further increases the RTT and complexity because the latency of a DDR memory is not constant and is sometimes more than 100 nanoseconds~\cite{DDRLatency}, whereas the latency of an embedded memory is much more stable and is usually a few nanoseconds.

Moreover, a shorter RTT also lowers the latency and bandwidth overhead introduced by re-transmission: a shorter RTT means quicker interaction between the sender and the receiver, and a shorter stalling time after a Frame error is detected. Therefore, for optimal efficiency, re-transmission should be implemented in a protocol layer where the RTT is minimized.

The RTT consists of two parts, the circuit delay (T\textsubscript{circuit}) and the cable delay (T\textsubscript{cable}). The circuit delay is the time the circuit logic spends to process and forward the data, including the latency introduced by the transceivers (T\textsubscript{gt}), the upper layer protocols (T\textsubscript{proto}), as well as the buffer queues (T\textsubscript{buffer}). The cable delay is the time the data travels along the cable, determined by the speed of light and the total link length. Assuming both directions of the link are symmetric, we have: 
\begin{equation} \label{e1}
RTT = 2 * (T_{\textit{circuit}} + T_{\textit{cable}})
\end{equation}
\begin{equation} \label{e2}
T_{\textit{circuit}} = T_{\textit{gt}} + T_{\textit{proto}} + T_{\textit{buffer}}
\end{equation}
\begin{equation} \label{e3}
T_{\textit{cable}} = \frac{L_{\textit{cable}}}{C}
\end{equation}
where {\em C} denotes the speed of light in the cable and $L_{\textit{cable}}$ denotes the link length.

While the T\textsubscript{cable} is a constant as the link length will not grow or shrink over time, the T\textsubscript{circuit} can vary in a very wide range, depending on the protocol layer where the RTT is measured. If re-transmission is implemented within or above the Network Layer, where more than two nodes are involved and the data needs to go across a switching node to be routed to the destination, then end-to-end RTT is used. Otherwise, if re-transmission is done hop-by-hop within the Link Layer, then hop-by-hop RTT is used.
\begin{figure}
    \centering
    \includegraphics[width=0.45\textwidth]{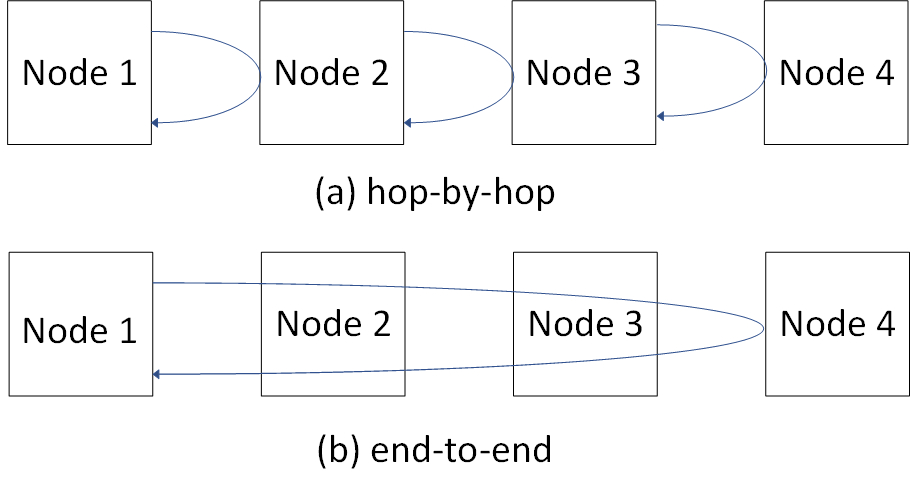}
    \caption{Hop-by-hop vs End-to-end}
    \label{retrans}
\end{figure}

Figure \ref{retrans} shows the difference between end-to-end and hop-by-hop. For end-to-end, the worst case RTT can be hundreds or thousands of times larger than the typical RTT. When the network is congested, the T\textsubscript{buffer} can be unpredictably large. Furthermore, congestion can cause frame losses, frame losses lead to re-transmission, and re-transmission can intensify network congestion, causing a positive feedback. For hop-by-hop, because there is no congestion at this level, the RTT will be constant and there will be no congestion-caused frame loss. Although end-to-end re-transmission is adopted by protocols such as TCP and Infiniband, according to the above discussion, hop-by-hop is better for minimizing the memory usage, the latency and the bandwidth overhead because it achieves the minimal RTT.

However, despite its significant advantages, re-transmission is seldom included in existing Link Layer protocols. One of the reasons we believe is related to the circuit area and complexity. The hardware implementation of a Link Layer protocol should not be heavy and power hungry. Specifically, a Link Layer protocol should not need megabytes of memory to function properly. In our case, assuming the line rate is 100 Gbps and the \textit{T}\textsubscript{\textit{Circuit}} is 100 nanoseconds, according to Equations \ref{e0}, \ref{e1} and \ref{e3} and the Constraint A, the S\textsubscript{retrans} required is no larger than 45 KB. The size is comparable to a CPU L1 cache, making Link Layer re-transmission feasible.

In conclusion, in a DCN, re-transmission should be done hop-by-hop within the Link Layer.

\subsection{Leveraging Hop-by-Hop Link Layer Re-Transmission}
Once hop-by-hop Link Layer re-transmission is chosen, a unique and vital property can be added to the constraint set, that is:
\begin{enumerate}[label=\Alph*]\bfseries
\setcounter{enumi}{2}
\item In the hop-by-hop Link Layer transmission, the receiver can assume that Frame N+1 will always arrive immediately after Frame N from the same sender.
\end{enumerate}

Such an assumption is not true for end-to-end transmission protocols such as any Ethernet-based protocol, where Frames from multiple senders can be routed to the same receiver. The receiver may receive Frame N and Frame N+1 from different sources. The traffic can also stop at Frame N if none of the senders continues to send data to the receiver after Frame N. However, for the Link Layer, a receiver is always paired to the same sender at the other end of the cable. If the user at the sender stops sending valid data after Frame N, the Link Layer protocol can pack invalid/idle data into Frames to create Frame N+1 and the subsequent Frames. The invalid Frames can be used by the protocol internally without being delivered to the user. This is an extremely useful property for hop-by-hop Link Layer re-transmission. We will discuss how it can be leveraged in the upcoming sections.

There is another equivalent expression of Constraint C that is worth emphasizing, i.e.:

\textbf{The receiver will never receive Frame N+1 before receiving Frame N} because in the hop-by-hop Link Layer transmission there is no buffer overflow caused by congestion. Starting from the sender logic, the data is handed over to the transceiver and then it is serialized, crosses the cable, is de-serialized, and finally it is handed over to the receiver logic. There can be a few bits that are not sampled, causing the link to be out-of-sync. However, there is no way that a whole Frame is lost during this process.

\subsection{ACK vs NACK}

ACK (acknowledgment) and NACK (negative acknowledgment) are the two possible acknowledgement mechanisms for re-transmission. For ACK, the receiver sends acknowledgements whenever it receives correct Frames. For NACK, the receiver sends acknowledgements whenever it receives Frames with bit errors. 

In a DCN context, NACKs have a significantly better efficiency over ACKs: let p denote the Frame Error Ratio (FER\footnote{\textbf{Frame Error Ratio}: ratio of Frames received with errors over total Frames received.}), and N denote the total number of Frames to be transmitted during a certain period. For ACK, at least N*(1-p) acknowledgments need to be transmitted from the receiver to the sender; For NACK, at least N*p negative acknowledgments are needed. In DCNs, as a result of Constraint B, p is much smaller than 1-p. Therefore, with NACKs, a much higher reverse channel bandwidth efficiency\footnote{\textbf{Bandwidth Efficiency}: ratio of the usable bandwidth to the line rate.} can be achieved compared to ACKs.

Nevertheless, for end-to-end re-transmission, reliability cannot be guaranteed with only NACKs and no ACKs. Assume Frame N is the last Frame to be transmitted from the sender to the receiver, and Frame N is dropped by an intermediate node (e.g., a switch). The receiver will never know that Frame N has been sent, hence no NACK will be generated. Similarly, the sender will never know that Frame N is not received, hence Frame N will not be re-transmitted. However, for hop-by-hop Link Layer re-transmission, with Constraint C, it is feasible to use only NACKs to achieve reliability, because there are always Frames being transmitted and none of them can be lost. They can only be corrupted. As a result, NACK is the acknowledgment mechanism we choose for RIFL.

\subsection{Summary}
In this section we have now provided the basis for RIFL. We summarize the characteristics here before describing its implementation:
\begin{itemize}
    \item Data corruption is handled by re-transmission.
    \item The buffers required by re-transmission can be implemented entirely using embedded memories.
    \item Link Layer frames will always arrive in sequence.
    \item We will use NACKs to reduce bandwidth overhead introduced by acknowledgments.
\end{itemize}

\section{Defining the RIFL Frames} \label{S3}
In Section \ref{S2}, we justified that Link Layer hop-by-hop re-transmission is an efficient solution for eliminating bit errors in DCNs. However, the protocol itself and its microarchitecture will also significantly impact the efficiency.

Without a concrete protocol, we are still far away from the final answer.

In this Section, we will define the RIFL Frames by answering the following questions:
\begin{enumerate}
    \item The Frame Structure: What are the header fields in a RIFL Frame?
    \item The Frame Size: How large is a Frame in RIFL?
\end{enumerate}

\subsection{High-Level Exploration of the Data Frame Structure} \label{S31}
There is no universal definition of {\em Frame}. In Section \ref{S2} we defined a {\em Frame} as the basic unit of data transmitted across the link. At higher protocol layers, we use the term {\em packet} to denote a bundle of data, such as an IP packet. A packet will be transmitted as a number of RIFL Link Layer Frames. To function properly, Link Layer Frames not only carry the payload, but also carry other essential signals. For example, when re-transmission or flow control events occur, the corresponding control signals need to be exchanged between the sender and the receiver. There should be Frames that carry such information. However, such events are assumed to occur much less frequently than regular data transmission. For bandwidth efficiency there is no reason to include both the control signals and the payload in every Frame.

We need to define different types of Frames. By functionality, we divide the Frames into the Data Frames and the Control Frames. The Data Frames are the Frames that carry the payload, and all the other Frames are Control Frames that help maintain state transitions. In a healthy link, most of the Frames being transmitted are Data Frames.

It is important to define the Data Frame structure well so that it serves the goal of making RIFL a low latency, high bandwidth, lightweight (small circuit area) and lossless Link Layer protocol. Section \ref{S2} showed that the circuit area is mainly impacted by the cable length and the microarchitecture of the protocol, and it is less relevant to the Data Frame structure. When define the Data Frame structure, we should mainly study its impact on the latency and the bandwidth efficiency.

\subsubsection{Header Fields}
To make the bandwidth overhead small, only essential information should be included in the header of the Data Frames.

First, to be able to detect any errors, a checksum must be included in every Data Frame. Second, a Data Frame should carry a Frame ID. Usually, there will be more than one Data Frame being transmitted during an RTT, so the Frame ID is used as the identifier to indicate which Data Frames should be re-transmitted when errors are detected. Third, for better granularity, a Data Frame should carry the information to indicate how many bytes in the payload are valid. Also, because any packet is divided into one or multiple Data Frames, there should be a marker in the Data Frame header to distinguish the end-of-packet Data Frames from other Data Frames, so that packet boundaries can be defined. Finally, for any Link Layer protocol, a line code should be adopted to re-align the data after deserialization. For 64b/66b encoding in Ethernet and Aurora~\cite{Aurora}, and 64b/67b encoding in Interlaken~\cite{INKL}, the encoding is done independently from the protocol framing. Different from the conventional protocols, in RIFL, to minimize the complexity and the latency, the line code is integrated into every Frame.

In summary, the Data Frame header should carry the following essential information: the checksum, the Frame ID, the count of valid bytes in the payload, the end-of-packet marker and the line code header.

\subsubsection{Data Frame Size}
The first decision RIFL made for the Data Frame size is to use a fixed frame size instead of a variable frame size. While a variable frame size is overall good for bandwidth efficiency, it is more complicated to implement, introduces longer latency, and requires a much larger buffer. Most importantly, a variable frame size introduces variable frame intervals (the difference of the arrival times between two adjacent frames), which can greatly increase the complexity of the re-transmission protocol. It is not worth sacrificing so much to save only three percent of the bandwidth. Thus, we only study the frame size impact of fixed size Data Frames. We start with exploring the impact of the Data Frame size on the bandwidth efficiency.

The following equation can yield the bandwidth efficiency:
\begin{equation} \label{eff}
\textit{Eff}_\textit{bandwidth} = (1-\frac{\textit{S}_\textit{Dheader}}{\textit{S}_\textit{DFrame}}) \times \textit{R}_\textit{DFrame}
\end{equation}
where \textit{Eff}\textsubscript{\textit{bandwidth}} denotes the bandwidth efficiency, S\textsubscript{\textit{Dheader}} denotes the size of the header in a Data Frame, S\textsubscript{\textit{DFrame}} denotes the Data Frame size, and \textit{R}\textsubscript{\textit{DFrame}} denotes the fraction of the Data Frames transmitted to all Frames transmitted.

By Constraint C, there are continuous Frames transmitted, regardless of whether there is valid data to transmit. Let \textit{R}\textsubscript{\textit{NDFrame}} denote the fraction of all the non-Data Frames, we get:
\begin{equation} \label{RNDF}
\textit{R}_\textit{DFrame} = 1 - \textit{R}_\textit{NDFrame}
\end{equation}

Assuming when an error is detected, on average, there are N\textsubscript{\textit{stall}} subsequent non-Data Frames (including the re-transmitted Data Frames and the Control Frames) being transmitted, we get:
\begin{equation} \label{RNDF2}
\textit{R}_\textit{NDFrame} = \textit{N}_\textit{stall} \times \textit{FER}
\end{equation}

Combining Equation \ref{eff}, \ref{RNDF}, \ref{RNDF2}, we get:
\begin{equation} \label{eff2}
\textit{Eff}_\textit{bandwidth} = (1-\frac{\textit{S}_\textit{Dheader}}{\textit{S}_\textit{DFrame}}) \times (1 - \textit{N}_\textit{stall} \times \textit{FER})
\end{equation}

where
\begin{equation} \label{fer}
\textit{FER} = 1 - (1 - \textit{BER})^{\textit{S}_\textit{DFrame}}
\end{equation}

According to Equation \ref{eff2}, a higher bandwidth efficiency is achieved by reducing the ratio of S\textsubscript{\textit{Dheader}} to S\textsubscript{\textit{DFrame}}, and minimizing N\textsubscript{\textit{stall}} and FER. Among the three factors, N\textsubscript{\textit{stall}} is mainly affected by the protocol design, while the others are mainly determined by S\textsubscript{\textit{DFrame}}.

For a good re-transmission protocol, most of the frames transmitted should be Data Frames. For environments with a low BER, R\textsubscript{\textit{NDFrame}} will be much smaller compared to the ratio of S\textsubscript{\textit{Dheader}} to S\textsubscript{\textit{DFrame}}. So, the \textit{Eff}\textsubscript{\textit{bandwidth}} will be mainly impacted by the ratio of S\textsubscript{\textit{Dheader}} to S\textsubscript{\textit{DFrame}}. As S\textsubscript{\textit{DFrame}} increases, S\textsubscript{\textit{Dheader}} will also increase because some of the header fields, such as the checksum, need to be expanded for a larger S\textsubscript{\textit{DFrame}}, but S\textsubscript{\textit{Dheader}} increases more slowly than S\textsubscript{\textit{DFrame}} increases. For example, among all the Cyclic Redundancy Check (CRC) codes that feature a Hamming Distance~\cite{hamming} (HD) of four (can detect at most three errors), 8-bit CRC codes can protect at most 119 bits of payload, while 16-bit CRC codes can protect at most 32751 bits of payload~\cite{CRC1}~\cite{CRCZOO}. Therefore, generally, the ratio of S\textsubscript{\textit{Dheader}} to S\textsubscript{\textit{DFrame}} decreases when S\textsubscript{\textit{DFrame}} increases. Nevertheless, this does not mean S\textsubscript{\textit{DFrame}} can be infinitely large. For the same BER, the larger S\textsubscript{\textit{DFrame}} is, the larger the FER is. Even though by Constraint B, the BER should be smaller than 10\textsuperscript{-7}, if S\textsubscript{\textit{DFrame}} is large enough, R\textsubscript{\textit{NDFrame}} can still impact \textit{Eff}\textsubscript{\textit{bandwidth}}.

In addition, a larger S\textsubscript{\textit{DFrame}} also means a larger latency. During transmission, the receiver can only verify the correctness of a Data Frame after all the bits of the Data Frame are received. To guarantee a lossless transmission, before examining the entire Data Frame, not a single bit of the Data Frame can be delivered from the receiver. That is to say, the larger S\textsubscript{\textit{DFrame}} is, the larger the latency will be introduced by checksum verification.

In summary, S\textsubscript{\textit{DFrame}} cannot be too small, otherwise the bandwidth overhead of the header will be too large. On the other hand, S\textsubscript{\textit{DFrame}} cannot be too large as well, otherwise the bandwidth can also be reduced because of a high FER, and the latency will also be too large.

\subsection{The Data Frame}
With the conclusions of Section \ref{S31}, we define the following Data Frame header fields:
\subsubsection{\textbf{Syncword (SYN)}}
SYN is a 2-bit line code header. It is also used as a marker to mark whether a Frame is a Data Frame or a Control Frame. Using the Verilog constant notation, in Data Frames, SYNs are set to 2'b01; in Control Frames, SYNs are set to 2'b10. A SYN of 2'b00 or 2'b11 is illegal, indicating that data is not aligned.
\subsubsection{\textbf{Payload}}
The user payload.
\subsubsection{\textbf{Meta Code}}
The Meta Code is used to indicate whether the Payload is not valid, partially valid, or all bytes of the Payload are valid. The end-of-packet marker is also encoded by the Meta Code. Table \ref{tab:Meta} shows the Meta Code encoding and the corresponding interpretation. With only two bits, the Meta Code cannot indicate how many bytes in the Payload are valid. It can only indicate whether all bytes of the Payload are valid. When not all bytes of the Payload are valid, the last byte of the Payload, which is certainly invalid as user data, becomes the Format Code.
\begin{table}
\centering
\caption{\label{tab:Meta}Meta Code Encoding} 
\begin{threeparttable}
\centering
 \begin{tabular}{|c c c c |}
 \hline
 Meta Code & Payload Valid & EOP & ABV \\ [0.4ex] 
 \hline
 00 & No & No & No\\ 
\hline
 01 & Yes & No & Yes\\ 
 \hline
 10 & Yes & Yes & Yes\\ 
 \hline
 11 & Yes & Yes & No\\ 
 \hline
\end{tabular}
\begin{tablenotes}
   \item[*] EOP: end of packet
   \item[*] ABV: all bytes valid
\end{tablenotes}
\end{threeparttable}
\end{table}

\subsubsection{\textbf{Format Code}}
The Format Code is an 8-bit field. It is used to indicate how many bytes in the Payload are valid when the Meta Code indicates that not all bytes of the Payload are valid. By combing the Meta Code and the Format Code, the count of the valid bytes in the payload and the end-of-packet marker mentioned in Section \ref{S31} can be represented with only a cost of two bits in the Data Frame header. Meanwhile, because the Format Code is limited to eight bits, it only works when the Payload size is not larger than 2048 bits (256 bytes).

\subsubsection{\textbf{Verification Code}} The Verification Code is the exclusive-or result of the checksum and the Frame ID. It combines the functionalities of the checksum and the Frame ID, i.e., the Verification Code is used to verify the correctness of the Data Frames as well as to locate the error Frame when an error is detected. More details of usage of the Verification Code will be illustrated in the next section.

\begin{figure*}
    \centering
    \includegraphics[scale=0.12]{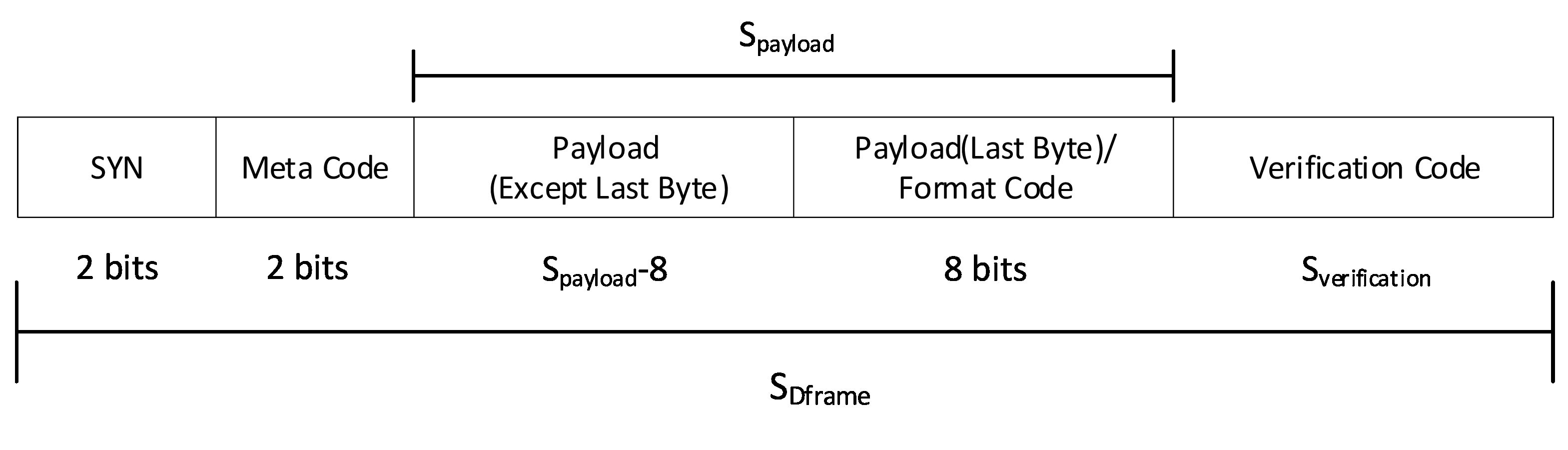}
    \caption{Data Frame Structure}
    \label{Format}
\end{figure*}

Figure~\ref{Format} shows the Data Frame Structure, where S\textsubscript{\textit{DFrame}}\footnote{We use bit as the size unit for the rest of this paper} denotes the Data Frame size, S\textsubscript{\textit{payload}} denotes the size of the Payload, and S\textsubscript{\textit{verification}} denotes the size of the Verification Code. We use Xilinx FPGAs for prototyping, and the available transceivers offer 32-bit, 64-bit and 128-bit interfaces ~\cite{GTH}~\cite{GTY}~\cite{GTM}. To minimize the latency and complexity of data width conversion, S\textsubscript{\textit{DFrame}} should be a multiple of the interface width of the transceiver. In our prototype, we set S\textsubscript{\textit{DFrame}} to be a power of two, and no less than 128. According to Figure~\ref{Format}, we get:
\begin{equation}\label{SDFrame}
\begin{aligned}
\textit{S}_\textit{DFrame} &=  \textit{S}_\textit{payload} + \textit{S}_\textit{Dheader} \\
&= \textit{S}_\textit{payload} + \textit{S}_\textit{verification} + 4
\end{aligned}
\end{equation}
If we assume \textit{R}\textsubscript{\textit{NDFrame}} is small, then \textit{R}\textsubscript{\textit{DFrame}} is close to 1. Combining Equation \ref{eff} and Equation \ref{SDFrame}, we get:
\begin{equation} \label{bandwidthefficiency}
\textit{Eff}_\textit{bandwidth} = 1-\frac{\textit{S}_\textit{verification}+4}{\textit{S}_\textit{DFrame}}
\end{equation}

As discussed in Section \ref{S31}, to minimize the latency and maximize the bandwidth efficiency, both S\textsubscript{\textit{DFrame}} and S\textsubscript{\textit{verification}} need to be small. Because S\textsubscript{\textit{DFrame}} is set to be a power of two, and no less than 128, and the Frame Code can only support up to 2048 bits of Payload, the S\textsubscript{\textit{DFrame}} options are limited to: 128, 256, 512, 1024, and 2048.

Let S\textsubscript{\textit{FrameID}} denote the size of the Frame ID field and S\textsubscript{\textit{checksum}} denote the size of the checksum. Because the Verification Code is the exclusive-or result of the checksum and the Frame ID, we get:
\begin{equation} \label{verification}
\textit{S}_\textit{verification} = \textit{Max}(\textit{S}_\textit{FrameID},\textit{S}_\textit{checksum})
\end{equation}
A valid tuple of (S\textsubscript{\textit{DFrame}},S\textsubscript{\textit{verification}}) should satisfy the following requirements:
\begin{enumerate}
    \item The size of the Frame ID should provide enough unique data frame IDs to cover all the data frames being sent during an RTT.
    \item For any BER that is better than 10\textsuperscript{-7}, the Mean Time Before Failure (MTBF)\footnote{In this paper, we define MTBF as the time to make the system failure possibility equal to 1\%} associated with the checksum should be at least longer than the lifetime of the circuit, say 100 years.
\end{enumerate}
The first requirement can be quantitatively described as:
\begin{equation} \label{Fid}
2^{\textit{S}_\textit{FrameID}} \geqslant \frac{\lambda_\textit{line} \times \textit{RTT}}{\textit{S}_\textit{DFrame}}
\end{equation}

The second requirement can be expressed by:
\begin{equation} \label{checksum}
(1-\textit{FFR})^{\frac{\lambda_\textit{line} \times \textit{MTBF}}{\textit{S}_\textit{DFrame}}} = 99\%
\end{equation}
where FFR denotes the Frame Failure Ratio, representing the ratio of the error Frames that cannot be detected by verifying the checksum to the total number of frames transmitted. In RIFL, we use a CRC code as the checksum. For an m-bit CRC code that features a Hamming Distance~\cite{hamming} (HD) of $n+1$, it can detect all error Frames that carry no more than n error bits. If the number of the error bits are more than n, one over 2\textsuperscript{m} of the error Frames cannot be detected. Therefore:
\begin{equation} \label{checksum1}
\textit{FFR} = \frac{1}{2^m} \times (1 - \sum_{i=0}^{n} \textit{P}(\textit{i}))
\end{equation}
where \textit{P}(\textit{i}) denotes the possibility of a frame carrying exactly i bits of errors:
\begin{equation} \label{pi}
\textit{P}(\textit{i}) = \binom{\textit{S}_\textit{DFrame}}{\textit{i}}\textit{BER}^\textit{i}(1-\textit{BER})^{\textit{S}_\textit{DFrame}-\textit{i}}
\end{equation}

There are a wide range of CRC codes listed in \cite{CRCZOO}. Let the line rate be 100 Gbps, RTT be 500 ns, and the BER be 10\textsuperscript{-7}. Combining Equations \ref{Fid}, \ref{checksum}, \ref{checksum1}, \ref{pi}, and the CRC codes in \cite{CRCZOO}, the minimal S\textsubscript{\textit{FrameID}} and S\textsubscript{\textit{checksum}} for different S\textsubscript{\textit{DFrame}} can be found in Table \ref{tab:verification}.
\begin{table}
\centering
\caption{\label{tab:verification}Minimal S\textsubscript{\textit{FrameID}} and S\textsubscript{\textit{checksum}} for different S\textsubscript{\textit{DFrame}}} 
\begin{threeparttable}
\centering
 \begin{tabular}{|c c c c c |}
 \hline
 S\textsubscript{\textit{DFrame}} & S\textsubscript{\textit{FrameID}} & S\textsubscript{\textit{checksum}} & HD & MTBF(year) \\ [0.4ex]  
 \hline
 128 & 9 & 8 & 4 & 9.7*10\textsuperscript{4}\\ 
\hline
 256 & 8 & 9 & 4 & 2.4*10\textsuperscript{4}\\ 
 \hline
 512 & 7 & 10 & 4 & 5.9*10\textsuperscript{3}\\ 
 \hline
 1024 & 6 & 11 & 4 & 1.5*10\textsuperscript{3}\\
 \hline
 2048 & 5 & 12 & 4 & 3.6*10\textsuperscript{2}\\
 \hline
\end{tabular}
\begin{tablenotes}
   \item[*] \textlambda\textsubscript{\textit{line}}: 100 Gbps\space\space\space
   \space\space\space RTT: 500 ns \space\space\space
   \space\space\space BER: 10\textsuperscript{-7}
\end{tablenotes}
\end{threeparttable}
\end{table}

Because the Payload is input from the user interface, and following the convention that the data width of the user interface should be a power of two, there should be a data width conversion module to convert the user input to the Payload. To minimize the latency and the complexity of the conversion module, the Payload should be byte-aligned:
\begin{equation} \label{payload}
\textit{S}_\textit{payload} \equiv 0 \mod 8
\end{equation}
Because we have limited S\textsubscript{\textit{DFrame}} to a power of two and to be no less than 128, we get:
\begin{equation} \label{sdof8}
\textit{S}_\textit{DFrame} \equiv 0 \mod 8
\end{equation}
Combining Equations \ref{SDFrame}, \ref{payload}, \ref{sdof8} , we get:
\begin{equation} \label{verification3}
\textit{S}_\textit{verification} \equiv 4 \mod 8
\end{equation}

The minimal S\textsubscript{\textit{verification}} and the corresponding \textit{Eff}\textsubscript{\textit{bandwidth}} for various values of \textit{S}\textsubscript{\textit{DFrame}} can be found in Table \ref{tab:verification2}.
\begin{table}
\centering
\caption{\label{tab:verification2}S\textsubscript{\textit{verification}} vs S\textsubscript{\textit{DFrame}}}
\centering
 \begin{tabular}{|c c c |}
 \hline
 S\textsubscript{\textit{DFrame}} & S\textsubscript{\textit{verification}} & \textit{Eff}\textsubscript{\textit{bandwidth}} \\ [0.4ex]  
 \hline
 128 & 12 & 87.5\%\\ 
\hline
 256 & 12 & 93.25\%\\ 
 \hline
 512 & 12 & 96.87\%\\ 
 \hline
 1024 & 12 & 98.43\%\\
 \hline
 2048 & 12 & 99.22\%\\
 \hline
\end{tabular}
\end{table}

Let 90\% be the acceptance threshold of the bandwidth efficiency, then the available options for the Data Frame size are 256, 512, 1024, and 2048 bits, and \textit{S}\textsubscript{\textit{verification}} should always be 12 bits. Because the \textit{S}\textsubscript{\textit{verification}} should be 12 bits, we extend the CRC code to 12 bits for a stronger protection. We choose not to extend the Frame ID field, because a larger \textit{S}\textsubscript{\textit{FrameID}} means larger \textit{S}\textsubscript{\textit{retrans}}, which leads to larger circuit area.

In summary, we defined the Data Frame fields and the size of each field in this subsection.

\subsection{The Control Frame}
As discussed in \ref{S31}, there should be Control Frames in RIFL to help maintain state transitions. Because the Control Frames are used much less than Data Frames, the size of the Control Frames does not have much affect on the protocol efficiency. Therefore, we don not need to further analyze the impact of the Control Frame size like we did for the Data Frame size. To minimize the complexity, the Control Frame size is set to be equal to the Data Frame size.

\begin{figure*}
    \centering
    \includegraphics[scale=0.12]{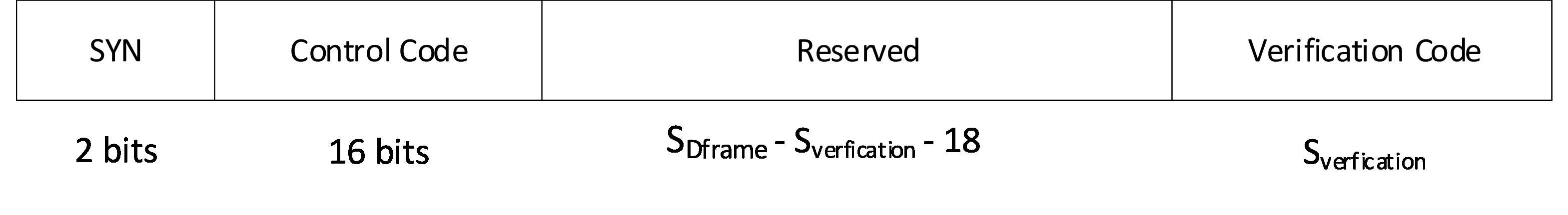}
    \caption{Control Frame Structure}
    \label{CFormat}
\end{figure*}

Figure \ref{CFormat} shows the Control Frame structure, where S\textsubscript{\textit{DFrame}} denotes the Data Frame size, S\textsubscript{\textit{verification}} denotes the size of the Verification Code.

The \textbf{SYN} and the \textbf{Verification Code} do the same thing in the Control Frames as they do in the Data Frames. The \textbf{Control Codes} are:

    \paragraph{\textbf{Idle}} This code indicates the sender is not in the normal data transfer state. This code is sent out when the sender is in the transition state between the pause state, the re-transmit state, and the normal state. Detailed explanations of each state will be introduced in the next section.
    \paragraph{\textbf{Pause Request}} This code is sent by the receiver when the link is out-of-sync. It notifies the sender to pause from sending data.
    \paragraph{\textbf{Re-transmit Request}} This code is sent by the receiver when a bad verification code is encountered. It tells the sender to switch from the normal data transmission to the re-transmission procedure.

\subsection{Summary}
In this section, we analyzed the Frame structure's impact on the bandwidth efficiency and the latency of the protocol. We defined the structure of the Data Frames and the Control Frames based on our analysis. It is worth noting that, the Frame sizes we chose are based on the interface data width of the transceivers we used for prototyping. For other types of transceivers that offer different interface data widths, the same analysis can be done again to determine the best Frame size options.

\section{Defining the RIFL Protocol} \label{S4}
In this Section, we will introduce how RIFL operates with the Frames we defined in Section \ref{S3}. By functionality, this section are divided as follows:
\begin{enumerate}
    \item \textbf{The TX and the RX Protocol}: How RIFL TX and RX side operates.
    \item \textbf{Re-transmission}: How re-transmission is done with the Verification Code we defined in the Section  \ref{S3}.
    \item \textbf{Flow Control} and \textbf{Clock Compensation}: Explanation of the flow control procedure and the clock compensation procedure.
    \item \textbf{Channel Bonding}: How RIFL aggregates multiple transceivers to achieve higher line rates.
\end{enumerate}

\subsection{The TX Protocol}\label{S41}
There are six states for the TX logic:
\begin{itemize}
    \item \textbf{Init}: In this state, invalid Data Frames are generated with Meta Code 2'b00, and Frame ID from zero to the max\footnote{The max value depends on how many bits are used for the Frame ID. E.g, if \textit{S}\textsubscript{\textit{FrameID}} is 8 bits, then the max value is 255. The \textit{S}\textsubscript{\textit{FrameID}} can be at most 12 bits because the \textit{S}\textsubscript{\textit{verification}} is set to 12 bits.}. The corresponding Verification Codes are also computed and inserted into each Frame. These invalid Data Frames will fill the re-transmission buffer during initialization. Throughout this state, the TX logic sends out back-to-back Pause Request Frames.
    \item \textbf{Send Pause}: Transmitting falls into this state when the RX logic detects that the link is out-of-sync, or right after the TX logic finishes initialization. Throughout this state, the TX logic sends out back-to-back Pause Request Frames.
    \item \textbf{Pause}: Transmitting falls into this state when Pause Requests Frames are received by the RX logic. Throughout this state, the TX logic sends back-to-back Idle Frames.
    \item \textbf{Retrans}: Transmitting falls into this state when Re-transmit Request Frames are received by the RX logic, or a re-transmission is resumed from an interruption caused by higher priority events. In this state, the TX logic can send three types of Frames: \textbf{Re-transmitted Data Frames}, \textbf{Idle Frames} or \textbf{Re-transmitted Request Frames}. More details will be elaborated in the upcoming Re-transmission subsection.
    \item \textbf{Send Retrans}: Transmitting falls into this state when an error is detected by the RX logic and there is no other higher priority condition. Throughout this state, the TX logic sends out back-to-back Re-transmit Request Frames.
    \item \textbf{Normal}: The normal data transmission state. As discussed previously, the link should stay in this state for most of the time if the BER is within the designed operation range (10\textsuperscript{-7} in our case). In this state, user is allowed to transmit data. When valid user data is input, the data is transformed into the Payload of one or multiple Data Frames. When user does not input valid data, invalid Data Frame with Meta Code 2'b00 are generated. In other words, in this state, the TX logic constantly sends out back-to-back Data Frames and copy them to the re-transmission buffer. Whenever user input is not valid, protocol-generated invalid Data Frames will be transmitted along the link to fill in the gaps.
\end{itemize}

\begin{figure*}
    \centering
    \includegraphics[scale=0.05]{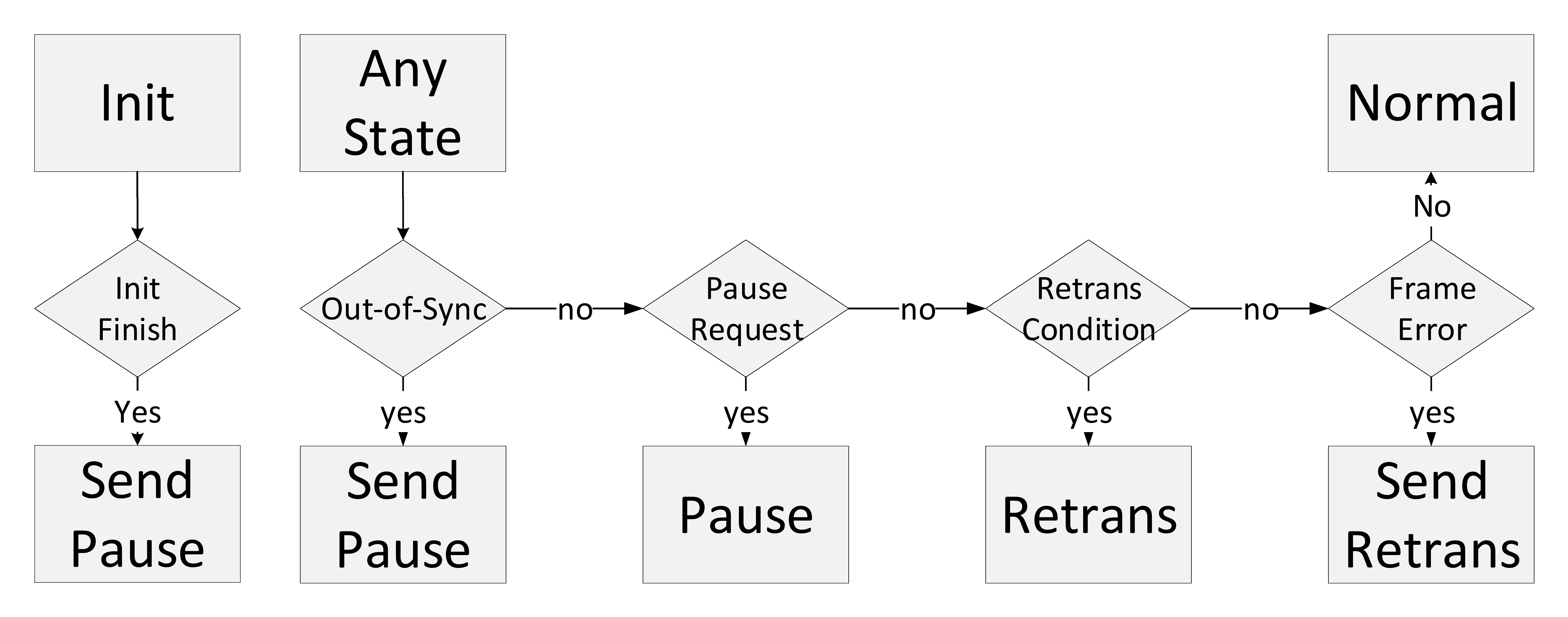}
    \caption{TX State Transition Diagram}
    \label{states}
\end{figure*}

Figure \ref{states} shows the state transition diagram for the TX logic. Except for the Init state, all the other states follow the same transition logic.

\subsection{The RX protocol}
There are in total five special events in RIFL: \textbf{Out-of-sync}, \textbf{Pause Request}, \textbf{Re-transmit Request}, \textbf{Frame Error} and \textbf{Flow Control}. The reactions of the TX logic to the first four events are already described in Sections~\ref{S41}. The Flow Control protocol will be introduced in Section~\ref{SFC}. The RX logic is responsible for monitoring such events and generating the event flags. Once an event is detected, the RX logic turns on the corresponding flag to notify the TX logic to make a proper reaction.

There is no state in the RX logic. All the special events are monitored independently and concurrently. The priority order of these events is presented in Figure \ref{states}. To prevent a Frame that carry errors from being recognized as a Control Frame, eight consecutive Pause Requests or Re-transmit Requests need to be received by the RX logic to activate the Pause or Re-transmit control flag. The Out-of-sync flag is turned on whenever an illegal \textbf{Syncword} is received. The Frame Error flag is turned on whenever a Data Frame with a wrong Verification Code is received.

\subsection{Re-transmission}\label{sec:retrans}
When both directions of the link are synchronized, the TX logic will switch between Normal, Retrans and Send Retrans states. The re-transmission falls into three scenarios:
\subsubsection{No error for both directions}
When there is no error for both directions of the link, both ends stay in the Normal state. In this scenario, the SYN of every Frame is always set to 2'b01 to represent a Data Frame. The Meta Code and the Payload are generated based on different scenarios of the user input. Every time a new Meta Code and Payload is generated, the 12-bit CRC checksum will be calculated. The Verification Code is then yielded by performing exclusive-or between the Frame ID and the checksum. After the TX logic sends out a Data Frame, the Frame ID will increment by one. Each Data Frame being sent out will also be copied to the re-transmission buffer. The re-transmission buffer is essentially a shift register, when a new entry is written, the oldest entry will be removed. Because \textit{S}\textsubscript{\textit{retrans}} is set to be equal to \textit{2}\textsuperscript{\textit{S}\textsubscript{\textit{FrameID}}}, each entry in the re-transmission buffer holds a Frame with an unique Frame ID. When a new Frame is written in to the buffer, the old Frame to be removed has the same Frame ID with the new Frame.

\subsubsection{Errors are detected in one of the directions}
When errors are detected in only one of the directions, the endpoint where the errors are detected enters the Send Retrans state, the other end enters the Retrans state. In the endpoint that is in the Send Retrans state, Frame Error flag is raised, its TX logic will send out back-to-back Re-transmit Request Frames. In the endpoint that is in the Retrans state, Re-transmit Request flag will be raised after the most recent received eight Control Frames are all Re-transmit Request. The TX logic will then perform the re-transmission procedure. Throughout the re-transmission procedure, the TX logic will send 2.5*\textit{2}\textsuperscript{\textit{S}\textsubscript{\textit{FrameID}}} Frames. The first 2*\textit{2}\textsuperscript{\textit{S}\textsubscript{\textit{FrameID}}} Frames are interleaved with Idle Frames and Re-transmitted Data Frames. The last 0.5*\textit{2}\textsuperscript{\textit{S}\textsubscript{\textit{FrameID}}} Frames are Idle Frames. After the last Frame of the re-transmission procedure is sent, if the Re-transmit Request flag is still raised, the TX logic will perform the re-transmission procedure all over again, until the Re-transmit Request Flag is down.

\subsubsection{Errors are detected in both directions}
When errors are detected in both direction, both endpoints will enter the Retrans state and start the re-transmission procedure. Different from the situation where only one direction detects the errors, for this scenario, the first 2*\textit{2}\textsuperscript{\textit{S}\textsubscript{\textit{FrameID}}} Frames will be Re-transmitted Data Frames interleaved with Idle Frames or Re-transmit Requests Frames. The last 0.5*\textit{2}\textsuperscript{\textit{S}\textsubscript{\textit{FrameID}}} Frames can also be either Idle Frames or Re-transmit Requests Frames. Whether to send the Re-transmit Requests Frames depends on the Frame Error flag is down or not.

By interleaving the Idle/Re-transmit Request Frames with the Re-transmitted Data Frames in the first 2*\textit{2}\textsuperscript{\textit{S}\textsubscript{\textit{FrameID}}} Frames, even there are errors in both direction, both endpoints can perform re-transmission while sending re-transmission notifications at the same time. In addition, when one of the endpoints stops sending the Re-transmit Request Frames, it will take a half of the RTT for the last Re-transmit Request Frame to arrive the other end, and only if the other end stops receiving the Re-transmit Request Frames, it can put down the Re-transmit Request flag. To cover this delay, the last 0.5*\textit{2}\textsuperscript{\textit{S}\textsubscript{\textit{FrameID}}} Frames are designed to be the buffer Frames.

Thus far, we have introduced the re-transmission procedure for the TX logic. On the RX side, there is also a procedure to verify if a Data Frame should be delivered to the user and if the Frame Error flag should be raised. Pseudo code of the verification procedure is shown in Listing 1.

\begin{lstlisting}[caption={RX Verification Procedure}]
Input: SYN, Meta Code, Payload, VCode
Output: Frame_Valid, Frame_Error
Init:
  FrameID = 0
  Threshold_FrameID = 16
Always:
  Checksum = CRC12({Meta Code,Payload})
  if VCode == FrameID ^ Checksum:
    if SYN == 2'b01:
      FrameID += 1
      if FrameID == Threshold_FrameID:
        Threshold_FrameID += 1
        Frame_Valid = True
        Frame_Error = False
      else:
        Frame_Valid = False
  else:
    FrameID = Threshold_FrameID - 16
    Frame_Valid = False
    Frame_Error = True
\end{lstlisting}

As shown in Listing 1, the RX logic keeps its own Frame ID counter ($FrameID$) and a threshold counter ($Threshold_{FrameID}$). $FrameID$ is initialized as 0 and $Threshold_{FrameID}$ is initialized as 16. When a Frame is received, the RX side will first calculate the CRC checksum of the Frame. The exclusive-or result of the checksum and $FrameID$ will then be compared against the Verification Code in the Frame. If the compare result is not equal, it implies the Frame has errors and the verification failed. The Frame Error flag will be raised and the Frame will not be delivered to the user. If the compare result is equal, meaning the verification passed, the RX logic will then examine the Syncword. If the Syncword is 2'b10, meaning the Frame is a Control Frame, the RX verification logic will not do anything. If the Frame is a Data Frame that carries a Syncword of 2'b01, $FrameID$ will then be compared against $Threshold_{FrameID}$, only if $FrameID$ is equal to $Threshold_{FrameID}$, the Data Frame can be delivered to the user, and both $FrameID$ and $Threshold_{FrameID}$ will then increment by one. If $FrameID$ is not equal to $Threshold_{FrameID}$, then only $FrameID$ will increment by one, the Frame will not be delivered to the user. In the case that the verification is failed. the $FrameID$ will be rolled back to $Threshold_{FrameID}$ minus 16.

The RX verification procedure is designed to deal with a special sequence of errors that can cause a false-positive verification result without the verification procedure. Here is an example of the special sequence of errors: assume Frame 68 has an error. A re-transmission is requested. Meanwhile, the subsequent frames, such as Frame 69 and Frame 70, are already on the fly. Because the 12-bit Verification Code is the exclusive-or result of the Frame ID and the CRC checksum, if either Frame 69 or Frame 70 has an error, they can be misrecognized as a correct Frame 68 - there is only one bit difference between the binary representations of 69 and 70 from 68. Also, because the TX logic will start re-transmission whenever the Re-transmit Request flag is raised, the re-transmission will not start exactly from Frame 68. Instead, it will start from a Frame sent before Frame 68. If some of the re-transmitted Frames before Frame 68 carry errors, they may also look like Frame 68 for the same reason. Thus, when an error is detected in Frame 68, the Frame ID will be rolled back to 52. We require the RX logic to see a correct sequence from Frame 52 to Frame 67 before accepting Frame 68. This means the RX logic must see a correct sequence of sixteen 12-bit Verification Codes. In this way, even a Frame with white noise (BER = 0.5) has only a chance of one over (2\textsuperscript{12})\textsuperscript{16} to be misrecognized as Frame 68. For BER better than 10\textsuperscript{-7}, the probability of a false-positive is even more negligible.

\subsection{Flow Control}\label{SFC}
As we discussed in Section~\ref{Introduction}, congestion control should be done at the Network Layer. However, besides congestion control, flow control is still necessary - the receiver may not be able to receive the data all the time, a method for the receiver to notify the sender to stop transmitting data is needed. To provide flow control, a buffer is added between the RX logic and the user interface. A simple ON/OFF flow control mechanism is adopted for low complexity. When the buffer queue length exceeds the ON threshold (\textit{Thr\textsubscript{\textit{ON}}}), the TX logic of the receiver will send out a flow control pause Frame\footnote{The flow control pause Frame is different from the Pause Request Control Frame}. When the buffer queue length drops below the OFF threshold (\textit{Thr\textsubscript{\textit{OFF}}}), the TX logic of the receiver will send out a flow control resume Frame. The sender completely stops transmitting any data after receiving the flow control pause Frame, and it resumes transmitting at the line rate after receiving the flow control resume Frame.

The size of the flow control buffer (\textit{S\textsubscript{\textit{FC}}}) must be carefully chosen to prevent any buffer overflow or starving during a flow control process - buffer overflow will cause frame losses and buffer starving will cause bandwidth under-utilization. Because it takes a half of the RTT for a flow control notification Frame to arrive from the receiver to the sender, during this period, the flow control buffer must reserve enough space to receive the Frames sent from the sender at the line rate, hence:
\begin{equation}\label{FC1}
\textit{S}_\textit{FC} - \textit{Thr}_\textit{ON} \geqslant \lambda_\textit{line} * \frac{\textit{RTT}}{2}
\end{equation}
Also, during this period, the buffer must also be able to deliver Frames to the user at the line rate, then we get:
\begin{equation}\label{FC2}
\textit{Thr}_\textit{OFF} \geqslant \lambda_\textit{line} * \frac{\textit{RTT}}{2}
\end{equation}
Lastly, \textit{Thr\textsubscript{\textit{ON}}} and \textit{Thr\textsubscript{\textit{OFF}}} must not be too close. Otherwise, frequently switching between ON and OFF will cause the flow control notification Frames occupying too much bandwidth of the reverse channel. For convenience, we set:
\begin{equation}\label{FC3}
\textit{Thr}_\textit{ON} - \textit{Thr}_\textit{OFF} \geqslant \lambda_\textit{line} * \frac{\textit{RTT}}{2}
\end{equation}
Combing Equation \ref{FC1} and \ref{FC2} and \ref{FC3}, we get:
\begin{equation}\label{FC5}
\textit{S}_\textit{FC} \geqslant \frac{3}{2} * \lambda_\textit{line} * \textit{RTT}
\end{equation}
and we can set:
\begin{equation}\label{FC6}
\textit{Thr}_\textit{ON} =  \frac{2}{3} * \textit{S}_\textit{FC}
\end{equation}
\begin{equation}\label{FC7}
\textit{Thr}_\textit{OFF} = \frac{1}{3} * \textit{S}_\textit{FC}
\end{equation}

After defining the flow control mechanism and the flow control buffer size, there is one remaining issue for flow control: bit error. Every Frame, including the flow control notification Frames, can end up being corrupted during transmitting. If there is a bit error in the flow control pause Frame, then it can result in a buffer overflow and a Frame loss. If there is a bit error in the flow control resume Frame, then the link may stop transmitting data forever. In our case, we extended the Meta Code encoding scheme and defined flow control notification Frames as special Data Frames. Previously, when the Meta Code is \textit{2'b00}, it indicates the Frame is an invalid Data Frame. Now, three types of Frames share Meta Code \textit{2'b00}. Only if the last byte of the Payload is \textit{2'h00}, it represents an invalid Data Frame. Otherwise, \textit{2'h01} represents a flow control pause Frame and \textit{2'h02} represents a flow control resume Frame.

By defining the flow control notification Frames as special Data Frames, the flow control notifications are guaranteed delivering to the sender. Even when there are bit errors, the flow control notifications will only be delayed, but not be missing. During the delay time, regular data transmissions at both sides of the link will be completely stopped because of re-transmission. Hence, there will be no data loss because of the flow control pause notifications not taking effect on time.

\subsection{Clock Compensation}\label{Compensation}
Although both sides of the link should operate at the same nominal line rate, the actual frequencies of their clocks will not be exactly the same because of the crystal oscillator frequency deviation. The endpoint with the faster clock will send data slightly faster than the slower end can receive. This will eventually overflow the slower end's receive buffer. With flow control, the issue can be resolved. However, it comes with a price of higher latency. Because flow control relies on the buffer queue length to slowly increase to \textit{Thr\textsubscript{\textit{ON}}} for a pause, the Frames at the end of the queue will experience large latency. It will be ideal if the TX logic at the faster endpoint can proactively regulate its rate. Because clocks are embedded into the data streams for serial transmission between transceivers, and RIFL directly interfaces with the transceivers, we are able to compare the frequency of the recovered clock with the frequency of the local clock to determine whether and when the TX logic should pause for one cycle for clock compensation. Details of the clock compensation implementation will be introduced in Section \ref{S5}.

\subsection{Channel Bonding}\label{ChannelBonding}
So far, we have introduced the single-lane protocol of RIFL. It works when both ends of the link only use a single transceiver for transmission. Nevertheless, although transceiver technology evolves rapidly, transceivers that support above 100 Gbps line rate are still rare to see. To achieve a bandwidth of hundreds gigabytes per second, channel bonding has to be done to aggregate the bandwidths of multiple transceivers. In RIFL, when multiple transceivers are used, every single pair of the transceivers runs the single-lane protocol. The channel bonding logic is responsible for dispatching the user data to each lane and aggregating the received data from each lane. To simplify the logic, we divide the user data into segments, the size of each segment is \textit{S}\textsubscript{\textit{Payload}}. At the TX side, the first segment goes to lane 1, the second goes to lane 2, and so on so forth. The same applies to the RX side, the Frame received from lane 1 is delivered first, followed by the Frame received from lane 2, and so on so forth. Because of the lane skew, lane 1 is not guaranteed to be the first lane to receive a Frame. The flow control buffer at each lane is used to overcome the lane skew. Details of the channel bonding implementation will be introduced in Section \ref{S5}.

\subsection{Summary}\label{Summary}
In this section, we have defined the RIFL protocols. We first introduced how TX and RX logic operates in general. We then added more details of re-transmission, flow control and clock compensation. Finally, we presented the channel bonding protocol.

More details on the implementation of the protocols are presented in Section~\ref{S5}.

\section{Implementation} \label{S5}
In this section, we present the FPGA implementation of RIFL that is open sourced at~\cite{RIFL}. RIFL is fully parameterized. Implementation options such as the Frame size and the transceiver line rate are exposed as synthesis parameters. For convenience, in this section, we demonstrate a four-lane implementation. In the implementation, each lane runs at 28 Gbps, and the Frame size is set to 256 bits.
\subsection{Top-Level Architecture}
\begin{figure}
    \centering
    \includegraphics[scale=0.08]{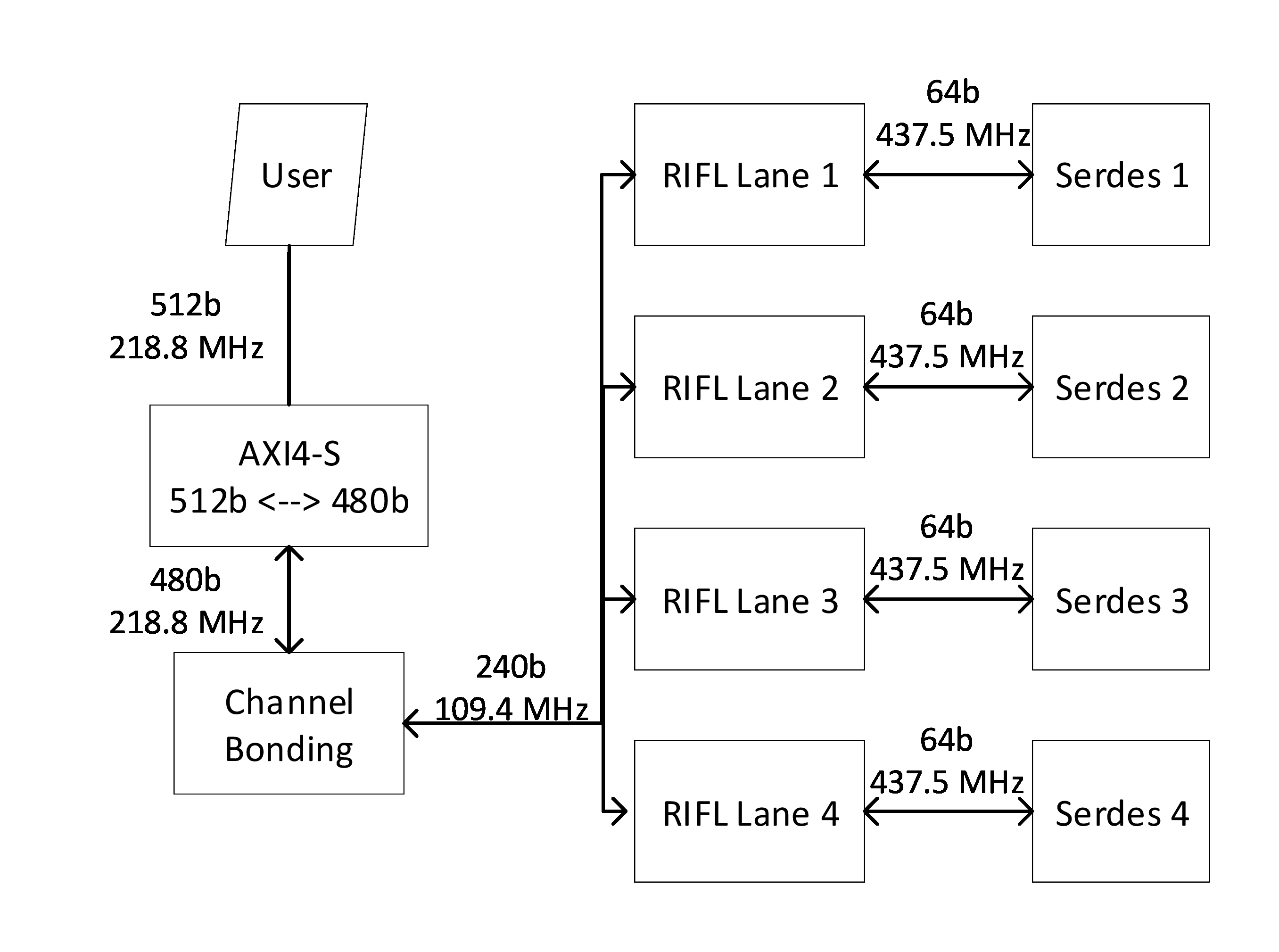}
    \caption{RIFL Top-Level Architecture}
    \label{MLane}
\end{figure}
Figure~\ref{MLane} shows the top-level architecture. RIFL provides a pair of AXI4-Stream~\cite{AXI4} interfaces to the user. Both interfaces consist of TDATA, TVALID, TKEEP, TLAST, and TREADY fields. With these fields, each flit~\footnote{\textbf{flit}: The data being transmitted in a single clock cycle} of the user data stream carries all the essential information we discussed in Section \ref{S31}. Adjacent to the user interfaces is the AXI4-Stream data width conversion block. It converts the stream width from any power of two to a multiple of \textit{S}\textsubscript{\textit{Payload}}.

When more than one transceiver is used, the AXI4-Stream data width converter will then be connected to the channel bonding module. In the TX path, the channel bonding module splits a single data stream into multiple data streams. In the RX path, it does the inverse. To provide more flexibility, two different channel bonding methods can be used in the channel bonding module: Temporal Channel Bonding and Spacial Channel Bonding. Temporal Channel Bonding splits a single data stream that runs at a higher frequency into multiple data streams that run at a lower frequency. After being split, the data width remains unchanged. Spacial Channel Bonding splits a single wider data stream into multiple narrower data streams and it does not change the frequency. In the example shown in Figure \ref{MLane}, both methods are used, the 512-bit AXI-4 Stream is first converted a 480-bit AXI-4 Stream. Then, inside of the channel bonding module, it is split into two 480-bit AXI-4 Streams running at half of the original frequency. Finally, each of the 480-bit AXI-4 Stream is split into two 240-bit streams. With two channel bonding methods, more user interface data width options are provided. For a four lane implementation with a Frame size of 256 bits, the data width can be 256 bits, 512 bits or 1024 bits. When implements RIFL on a low speed device such as a low end FPGA, wider interfaces with lower frequency can help timing closure, while on a high speed device, narrower interfaces are ideal for smaller circuit area.

If there is only a single lane, then the channel bonding module will be omitted. The AXI4-Stream data width converter will directly connect the single-lane logic. Details of the single-lane architecture will be presented in the next subsection.

\subsection{Single-Lane Architecture}
\begin{figure*}
    \centering
    \includegraphics[scale=0.09]{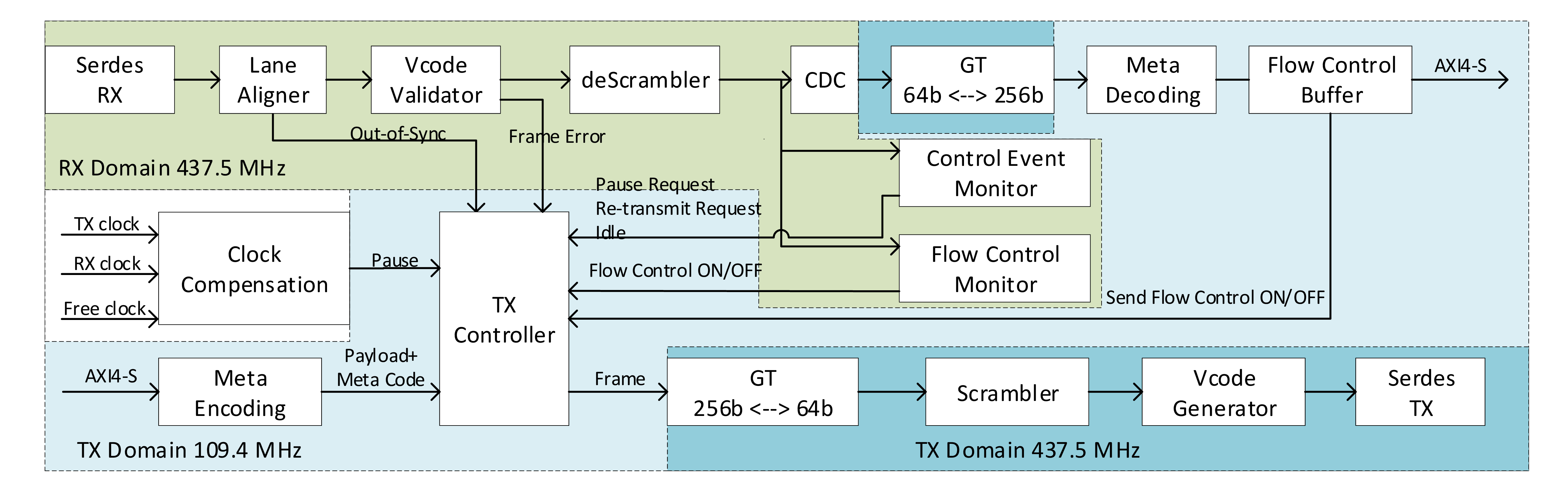}
    \caption{RIFL Single-Lane Architecture}
    \label{SLane}
\end{figure*}
Figure \ref{SLane} shows the single-lane architecture of RIFL. As shown in the figure, there are two clock domains: the RX Domain is driven by the recovered clock generated by the transceiver, and the TX Domain is driven by two local clocks - a high speed clock drives transceiver-facing logic and a low speed clock drives the rest of the protocol logic. The high and low speed clocks are derived from the same clock source. The frequency of the faster one is a power of two times of the frequency of the slower one. Hence, the two TX clocks are synchronous to each other. In the example, the high speed clock runs at 437.5 MHz and the lower speed clock runs at 109.4 MHz.

In the RX domain, the Lane Aligner converts the unaligned transceiver output stream to an aligned stream by locating the position of the Syncword. The Lane Aligner is essentially a two-level cascaded multiplexer chain. After the Lane Aligner, the Verification Code Validator is used to verify the correctness of the Verification Code. It is responsible for raising the Frame Error Flag. The scrambler and the descrambler used in RIFL are implemented in linear-feedback shift registers (LFSRs). The standard 33-bit scrambler code $(1+x^{13}+x^{33}$) is adopted for good DC balance and transition density ~\cite{scrambler}. After descrambling, the Clock Domain Crossing (CDC) module filters out the non-Data Frames by checking the Syncword. It then converts the filtered stream from the RX Domain to the TX Domain using a low latency asynchronous FIFO. The Control Event Monitor and the Flow Control Monitor are responsible for checking every Frame and generating the Pause Request flag, the Re-transmit flag and the flow control ON/OFF notifications. 

In the TX domain, the modules that are closer to the transceiver are driven by the high speed clock. They are the scrambler and the Verification Code generator. A pair of the GT data width converters are used to perform the conversion between the high-speed narrow stream used by the transceiver and the low-speed wide stream used by the rest of the protocol logic. The modules driven by the low-speed clock are the TX Controller, the Meta Code Encoding and Decoding modules, and the Flow Control Buffer. The finite-state machine (FSM) in the TX Controller implements the TX logic described in Section \ref{S41}. The Meta Code Encoding and Decoding modules convert the AXI4-Stream signals to the Meta Code signals. The Flow Control Buffer is a synchronous FIFO. It monitors its buffer queue length and issues flow control requests to the TX Controller.

Finally, the Clock Compensation module takes the TX clock and the RX clock from the transceiver, and a free-running clock as inputs. Each transceiver clock drives a gray code counter.  Both counters are then brought to the free-running clock domain for comparison. If the counter of the TX clock increases faster than the counter of the RX clock, then the difference of the counter values will be kept in a register. Whenever the difference increases, the Clock Compensation module will issue N cycles of pause signals to the TX controller. N is equal to the change of the difference between comparisons.

\section{Performance Evaluation} \label{S6}
We have validated the functional correctness of RIFL on both Intel and Xilinx devices for line rates from 25~Gbps to 200~Gbps. In this section, we present the performance results of RIFL that we obtained from Xilinx devices. We will first introduce our test setup. Then, we will compare the bandwidth efficiency, the latency and the resource usage between RIFL and Xilinx's Aurora~\cite{Aurora}, Interlaken~\cite{Xilinx_INKL} and 100G Ethernet (CMAC)~\cite{Xilinx_CMAC} implementations. We will then provide RIFL's performance result under various BER to demonstrate RIFL's reliability.

\subsection{Experimental setup}
\begin{figure*}
    \centering
     \begin{subfigure}[b]{0.4\textwidth}
         \centering
         \includegraphics[width=\textwidth]{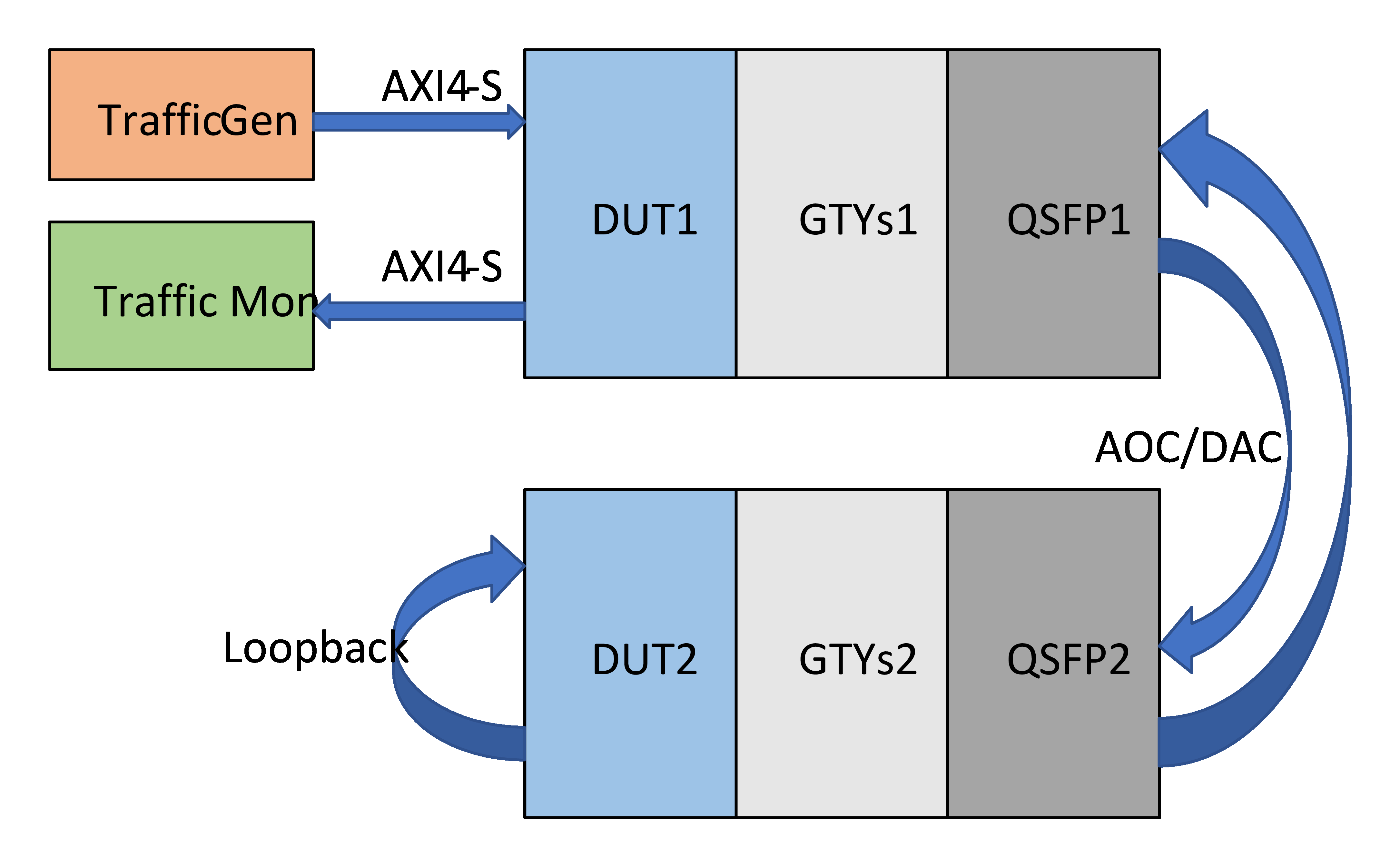}
         \caption{Performance Comparison Test Setup}
         \label{comparesetup}
     \end{subfigure}
     \begin{subfigure}[b]{0.4\textwidth}
         \centering
         \includegraphics[width=\textwidth]{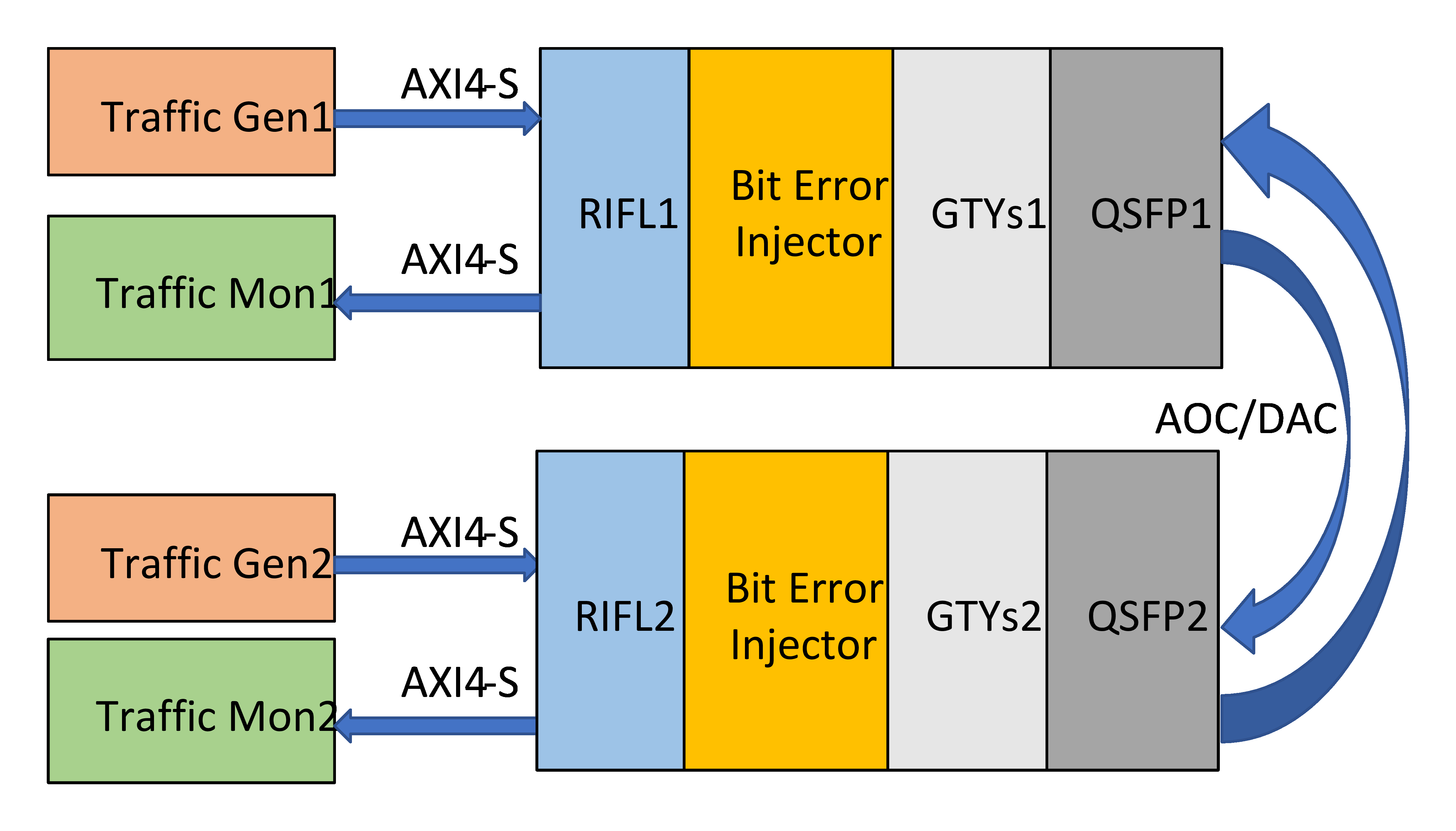}
         \caption{Reliability Test Setup}
         \label{bersetup}
     \end{subfigure}
     \caption{Performance Test Setups}
\end{figure*}
Our prototype is implemented on Fidus Siderwinder-100 (SW100)~\cite{sw100} boards. There are two QSFP28 ports on the board, connected to an XCZU19EG FPGA. Ten-meter Active Optical Cable (AOC) and 3-meter Direct Attach Copper (DAC) cables are used for the QSFP28 connections. For the sake of simplicity, we only present the results for the AOC in this Section.

A software-defined AXI4-Stream traffic generator is built to generate the testing traffic. This traffic generator allows AXI4-Stream traffic to be defined cycle by cycle in CSV format. The CSV file is then encoded into binary format and moved from an X86/ARM host to the FPGA memory. The hardware driver of the traffic generator retrieves the traffic data from the FPGA memory, performs decoding, and generates the traffic in a cycle-accurate manner according to the CSV definition.

A traffic validator is also built. It can cache the transmitted packets and compare them against the loopback traffic to verify the correctness. It also internally time-stamps each packet to monitor the bandwidth and latency.

Two different tests are designed for the performance comparison and the reliability test. The setup shown in Figure \ref{comparesetup} is used for performance comparison between the RIFL implementations and the Xilinx cores. The designs under test (DUTs) are placed in two FPGA boards to represent their general use case. The bandwidth efficiency and the RTT is measured in the first board. The point-to-point latency is yielded by halving the RTT - assuming the latencies for both direction are the same. For fair comparison, all DUTs use four Xilinx GTY transceivers. The line rate of each transceiver is set to 25.78 Gbps.

The reliability test setup is shown in Figure \ref{bersetup}. In this test, the same BER is imposed to both directions. To make the error patterns of the two directions independent, their random seeds are set different. In this case, the point-to-point latency cannot be considered as a half of the RTT anymore, because the link is not symmetric. For example, in a round trip, errors may happen in one of the directions, causing the latencies of both directions to be unequal. Therefore, the point-to-point latency has to be directly measured. As a result, both RIFL cores are placed in the same FPGA. Traffic generators and traffic validators are connected to both RIFL cores. The bandwidth efficiency and the average latency are computed by averaging the test results of both directions. The tail latencies are computed from the aggregated results of both directions. In the reliability test, each RIFL core uses four GTYs~\cite{GTY} running at 28 Gbps. The aggregated line rate is 112 Gbps, which is the maximum line rate a QSFP28 cable can support.

\subsection{RIFL vs Aurora vs Interlaken vs CMAC}
\begin{figure*}
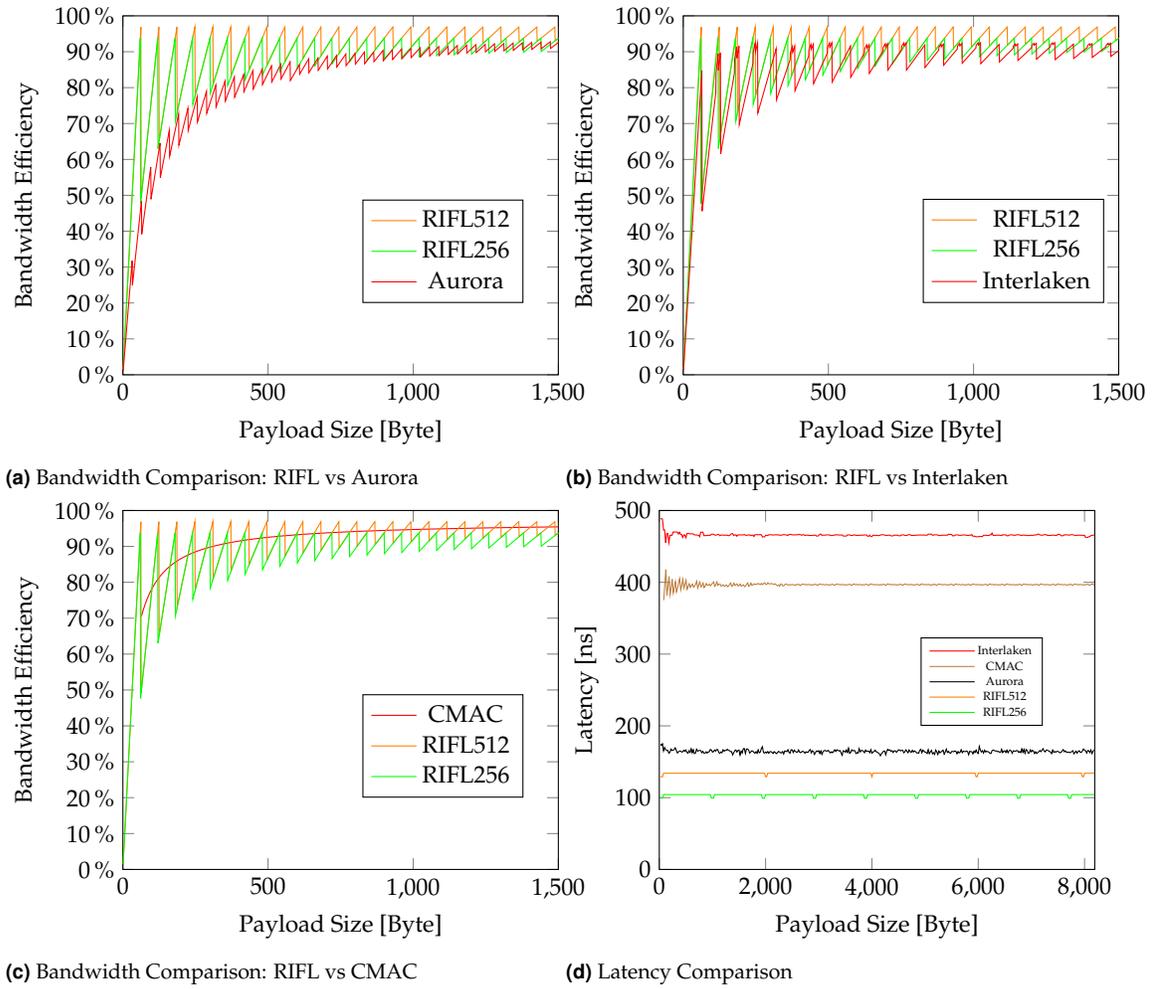

\centering
\begin{subfigure}[b]{0.4\textwidth}

\caption{Bandwidth Comparison: RIFL vs Aurora}
\label{bandwidth1}
\end{subfigure}
\begin{subfigure}[b]{0.4\textwidth}
%
\caption{Bandwidth Comparison: RIFL vs Interlaken}
\label{bandwidth2}
\end{subfigure}
\begin{subfigure}[b]{0.4\textwidth}
%
\caption{Bandwidth Comparison: RIFL vs CMAC}
\label{bandwidth3}
\end{subfigure}
\begin{subfigure}[b]{0.4\textwidth}
%
\caption{Latency Comparison}
\label{latency}
\end{subfigure}
\caption{Performance Comparison between RIFL, Aurora, Interlaken and CMAC}
\end{figure*}

\begin{figure*}
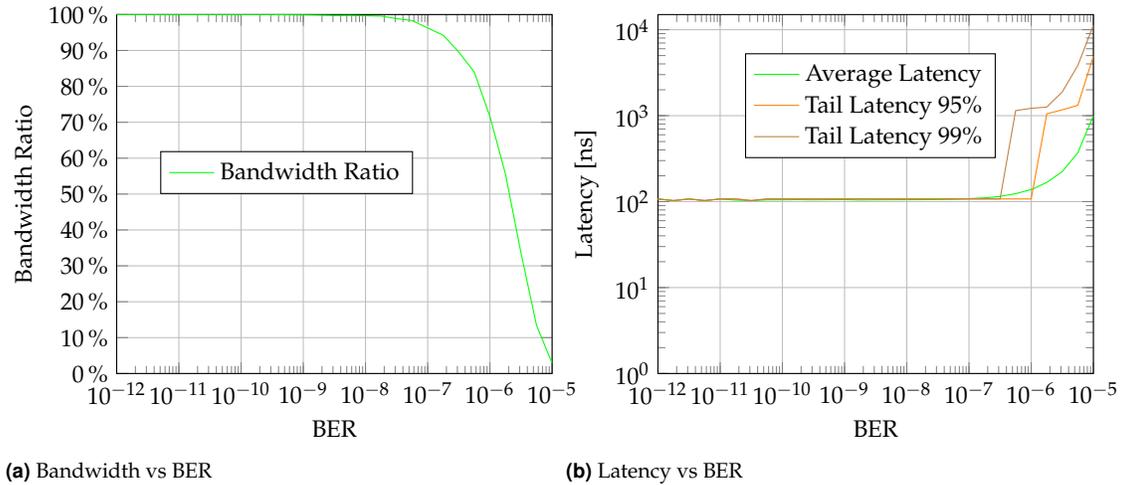

\centering
\begin{subfigure}[b]{0.4\textwidth}
%
\caption{Bandwidth vs BER}
\label{BERbandwidth}
\end{subfigure}
\begin{subfigure}[b]{0.4\textwidth}
%
\caption{Latency vs BER}
\label{BERlatency}
\end{subfigure}
\caption{Bandwidth and Latency under different BERs}
\end{figure*}

In this subsection, we compare the bandwidth efficiency, latency, and resource usage performance between RIFL, Aurora, Interlaken and CMAC.

For bandwidth efficiency comparison, we test the bandwidth efficiency results for different user payload sizes. The payload sizes sweep from 1 byte to 8192 bytes\footnote{CMAC starts at 64 bytes because its minimal accepted payload size is 64 bytes.}, with a step of one byte. When the size of a payload is larger than the maximum frame size of the DUT (32 bytes for RIFL256, 64 bytes for RIFL512, Interlaken and Aurora, 9600 bytes for CMAC), it is divided into multiple frames for transmission. For each payload size, a traffic of ten gigabytes is sent. The traffic generator saturates the available bandwidth of the DUT by sending out a flit of traffic whenever the DUT can accept one.

Figures \ref{bandwidth1}, \ref{bandwidth2} and \ref{bandwidth3} show the bandwidth efficiency comparison between RIFL, Aurora, Interlaken and CMAC. In the figures, RIFL256 represents the RIFL implementation with a Frame size of 256 bits and RIFL 512 represents RIFL with a Frame size of 512 bits. To preserve more details for small payload sizes, the results for payload sizes that are larger than 1500 bytes are not included in the figures. As the figures show, in terms of bandwidth efficiency, from the best to the worst, it is CMAC, RIFL512, RIFL256 and Interlaken. Unlike the zigzag curves of the other three cores, CMAC shows a much smoother curve. It is because for RIFL, Aurora and Interlaken, if the payload size is not a multiple of the user interface data width, then for the last flit of the packet, only a fraction of the user interface will receive valid data. After receiving the partial valid flit, the entire flit is fed into the pipeline, the invalid bits are replaced with bubbles. Meanwhile, for CMAC, the data received from the user interface is first buffered, and is then reconstructed. The last flit of packet N can be concatenated with the first flit of packet N+1 to eliminate the pipeline bubbles as much as possible. While buffering and reconstructing benefit the bandwidth efficiency, they come with a trade-off of the latency and the complexity.

For latency comparison, the same traffic patterns are used. Same with the bandwidth comparison, the traffic generator saturates the available bandwidth of the DUT.

Figure \ref{latency} shows the point-to-point latency comparison result. From the best to the worst, it is RIFL256, RIFL512, Aurora, CMAC and Interlaken. For CMAC, as previously mentioned, by buffering and reconstructing the user packets, the latency is increased. The latency for small packets varies significantly more than the large ones. For Aurora and Interlaken, without knowing their implementation details, we cannot infer what form up their latency. However, we are confident that it is our micro-architecture optimizations mentioned in the previous sections that make RIFLs the lowest latency implementations.

Table \ref{tab:resource}
\begin{table}
\caption{\label{tab:resource}Resource Comparison} 
\centering
\resizebox{0.9\columnwidth}{!}{
 \begin{tabular}{|l c c c c|}
 \hline
 Protocol & LUTs & Flip Flops & BRAM36Ks & DSPs \\ [0.4ex] 
 \hline
 RIFL(256,256) & 15308 & 15935 & 16 & 0\\ 
 \hline
 RIFL(256,1024) & 20048 & 14098 & 16 & 0\\
 \hline
 RIFL(512,512) & 28995 & 28960 & 32 & 0\\
 \hline
 Aurora & 10192 & 9447 & 4 & 0\\
\hline 
\end{tabular}
}
\end{table}
shows the resource usage comparison between three different implementations of RIFL and Aurora. In Table \ref{tab:resource}, RIFL(X,Y) represents RIFL with a Frame size of X bits and a user interface width of Y bits. Interlaken and Aurora are not included in the resource usage comparison because they are both hard cores, i.e., they are not implemented in FPGA soft logic.

It can be learned from the table that RIFL uses more resources than Aurora. One of the main reasons is that RIFL adds the re-transmission buffer and the flow control buffer for reliability. Another reason is that our FPGA prototype is not fully optimized for resource usage. For example, the data widths of BRAMs in the Sidewinder board are at most 64 bits while the buffer data width in RIFL is equal to its Frame size, being at least 256 bits. Although the capacity of a single BRAM is enough for the flow control buffer, we have to use multiple BRAMs for enough data width. Both reasons are related to the FPGA itself. If RIFL is hardened, the resource usage can be significantly reduced.

\subsection{Reliability Test}
\begin{table}
\centering
\caption{\label{tab:MTBF}$MTBF$ vs $BER$}
\resizebox{0.4\columnwidth}{!}{
\centering
 \begin{tabular}{|c | c|}
 \hline
 $BER$ & $MTBF$ (year) \\ [0.4ex]  
 \hline
 1.00E-11 & 1.81E+23\\ 
 \hline
 1.00E-09 & 1.81E+15\\
 \hline
 1.00E-07 & 1.88E+8\\
 \hline
 1.00E-05 & 6.33\\
 \hline
\end{tabular}
}
\end{table}

In this subsection, we present the bandwidth ratio, latency, and MTBF result of RIFL256 under different BERs. The bandwidth ratio is the ratio of the bandwidth under current BER to the bandwidth of a error-free link.

In the test, the size of the traffic is set to ten gigabytes. The traffic consists of mixed length packets. Payload sizes are randomly distributed from 1 byte to 8192 bytes. The BERs sweep from $10^{-12}$ to $10^{-5}$, with a step of $10^{\,0.25}$.

As shown in Figures \ref{BERbandwidth} and \ref{BERlatency}, the bandwidth and latency of RIFL do not degrade until the BER increases beyond about $10^{-7}$. The bandwidth ratio starts to drop when the BER is $5.6 \times 10^{-10}$, and it drops to $96.3\%$ when the BER is $10^{-7}$. The results agree with the theoretical calculation result of Equation \ref{bandwidthefficiency}. 

The latency of RIFL starts to increase when the BER is worse than $1.7 \times 10^{-6}$. When the BER is better than $10^{-7}$, the average latency and the tail latencies remain within 107 nanoseconds. This also agrees with the theoretical calculation.

As we discussed in Section \ref{sec:retrans}, during a re-transmisson, even a Frame of white noise is impossible to be mis-detected as a correct Frame. Therefore, for RIFL, Equation \ref{checksum} should be modified as:
\begin{equation}
(1-\textit{FFR})^{\frac{\lambda_\textit{actual} \times \textit{MTBF}}{\textit{S}_\textit{DFrame}}} = 99\%
\end{equation}
where \textlambda\textsubscript{\textit{actual}} denotes the actual bandwidth. With the bandwidth result, MTBF can be calculated.

As shown in Table~\ref{tab:MTBF}, when BER is $10^{-7}$, the MTBF is $1.88 \times 10^{7}$ years. Therefore, it is safe to claim that RIFL is reliable for any BER that is better than $10^{-7}$.

\subsection{Cross-Vender Communication}
We have successfully validated RIFL on a link between an Intel Agilex device and a Xilinx Vertex Ultrascale+ device.

\subsection{Summary}
In this section, we compare the latency and bandwidth efficiency result between two implementations of RIFL and three other Link Layer protocol implementations. We show that RIFL has the best latency and second best bandwidth efficiency while it is the only protocol that ensures lossless transmission. We also show RIFL can keep good performance and long MTBF when the BER is better than \textit{10\textsuperscript{\textit{-7}}}.

\section{Related Work} \label{S7}
In this section we describe the works that are most relevant to RIFL.

Ethernet~\cite{IEEE_Eth} was introduced in the 1980s and it is the most common protocol used in modern data centers~\cite{top500}. In the three-layer model we introduced in Section \ref{Introduction}, Ethernet includes not only Layer 1 functionalities, but also some Layer 2 functionalities, such as switching. Ethernet (Layer 1) allows variable frame sizes from 72 bytes to 1530 bytes (some implementations allow jumbo frames larger than 9000 bytes, but it is not compatible with the IEEE 802.3 standard). A 32-bit CRC is included in every Ethernet frame, enabling error detection but not error correction. Any re-transmission protocol working on top of Ethernet has to be end-to-end, which means Constraint C is not met any more. Moreover, the re-transmission buffer has to be large enough to handle a burst of the maximum-size frames. To summarize, a re-transmission protocol working on top of Ethernet would be more complex and less efficient than RIFL. Also, the experimental results in Section \ref{S6} show that RIFL perform better than CMAC, which is the Xilinx 100G Ethernet implementation~\cite{Xilinx_CMAC}.

Aurora \cite{Aurora} is a link layer protocol developed by Xilinx. It is made for point-to-point communication between FPGAs. There are two versions of Aurora, using two different line codes: 8b/10b for lower line rates and 64b/66b for higher line rates. The user payload is broken into multiple eight-byte frames called Data Blocks. The remaining bytes are transmitted using a special frame called the Separator Block. The Separator Block serves as an indicator of the end of a packet. A 32-bit CRC code is used in Aurora for error detection. Flow control directives are also provided.

Interlaken \cite{INKL} is invented by Cisco Systems and Cortina Systems. It uses 64b/67b encoding for better DC balance. There are two methods of packetization for Interlaken: BurstMax and BurstShort. The user payload is first broken into multiple 64-byte blocks and then transmitted using the BurstMax method. The remaining bytes are transmitted using BurstShort. The size of BurstShort can be from 32 bytes to 56 bytes, with 8-byte increments. Both BurstMax and BurstShort are ended with an 8-byte block named the Control Word. A 24-bit CRC code is integrated into the Control Word. Interlaken also provides in-band and out-of-band flow control, as well as out-of-band re-transmission.

Sanchez Correa et al. \cite{micro} create a protocol stack for FPGA-based high performance computing. Their Layer 1 is based on the 10 Gigabit Media Independent Interface (XGMII), limiting the throughput per lane to 10 Gbps. Their work is based on the assumption that the link channels are error free, hence reliability is not being taken care of at all.

None of the related works described here can provide or implement the low-latency, high bandwidth and especially reliable protocol that we require for our Layer 1 link layer protocol.

\section{Conclusion} \label{S8}
We have presented RIFL, a low latency and reliable Link Layer network protocol. Because of its novel in-band re-transmission protocol, RIFL is capable of providing lossless point-to-point links with ultra-low latency and high bandwidth. We implemented RIFL on Sidewinder boards and showed that at the line rate of 112 Gbps, approximately 100 nanoseconds point-to-point latency is achieved. We have also demonstrated that RIFL is capable of correcting all the data corruptions for standard point-to-point links.

With RIFL at the bottom layer, there is no need for the upper layer protocols to deal with any checksum. Therefore, the logic of the upper layer protocols can be simplified, and more resources can be used to deal with congestion control. This suggests that it is feasible to build a low-latency, high-bandwidth network for a data center environment based on RIFL. Our future work will address the Network Layer to enable congestion-free multi-hop communication.

\section{Acknowledgements}
This work is generously supported by Xilinx, Alibaba and NSERC. The authors declare that there is no conflict of interest regarding the publication of this paper.

\bibliography{sample}
\end{document}